# Identity-enabled CDMA LiDAR for massively parallel ranging with a single-element receiver

Yixiu Shen[1], Zi Heng Lim[1], Guangya Zhou[1*]

Light detection and ranging (LiDAR) have emerged as a crucial tool for high-resolution 3D imaging, particularly in autonomous vehicles, remote sensing, and augmented reality. However, the increasing demand for faster acquisition speed and higher resolution in LiDAR systems has highlighted the limitations of traditional mechanical scanning methods. This study introduces a novel wavelength-multiplexed code-division multiple access (CDMA) parallel laser ranging approach with a single-pixel receiver to address these challenges. By leveraging the unique properties of Gold-sequences in a direct-sequence spread spectrum (DSSS) framework, our design enables comprehensive parallelization in detection and ranging activities to significantly enhance system efficiency and user capacity. The proposed coaxial architecture simplifies hardware requirements using a single avalanche photodiode (APD) for multi-reception, reducing susceptibility to ambient noise and external interferences. We demonstrate 3D imaging at 5 m and 10 m, and the experimental results highlight the capability of our CDMA LiDAR system to achieve 40 parallel ranging channels with centimeter-level depth resolution and an angular resolution of 0.03°. Furthermore, our system allows for user identification modulation, enabling identity-based ranging among different users. The robustness of our proposed system against interference and speckle noise and near-far signal problems, combined with its potential for miniaturization and integration into chip-scale optics, presents a promising avenue to develop high-performance, compact LiDAR systems suitable for commercial applications.

Light detection and ranging (LiDAR) is rapidly becoming pervasive due to its exceptional imaging efficiency, high-resolution data capture, and impressive detection range. This technology generates swift and accurate 3D point cloud data, effectively overcoming the intrinsic limitations of traditional camera vision, especially their dependency on environmental lighting conditions [1, 2]. As a result, LiDAR now offers an expanded field-of-view (FOV) and rich multi-dimensional insights, pivotal in advancing autonomous vehicles, remote sensing, and augmented reality [3-5]. To further enhance acquisition speed and resolution while addressing the challenges of performance degradation in ultra-high-speed mechanical scanning [6], the adoption of parallelization strategies is a critical focus in the evolution of LiDAR technology.

Existing modulation techniques in massively parallel LiDARs encounter temporal and frequency congestion during simultaneous multi-ranging activities [7-9]. Inspired by multiple access technologies in digital communications, strategies such as block partitioning are implemented to counteract mutual interference. Despite utilizing techniques like frequency hopping (FH), frequency-division multiplexing (FDM), and wavelength-division multiplexing (WDM) to subdivide into sub-frequency bands for multi-channel measurements [1, 10, 11], an increase in channel count invariably compromises precision and certainty due to the narrow bandwidth. Furthermore, limited bandwidth causes inevitable frequency overlaps and increases susceptibility to signal jamming. Significant strides have been made in commercial pulsed time-of-flight (ToF) and continuous-wave (CW) LiDARs that employ space-division multiplexing (SDM) [12, 13] and time-division multiplexing (TDM) [14]. Nevertheless, these systems grapple with stringent demand in signal-to-noise ratio (SNR) and the escalated costs associated with detectors in complex multiple-input and multiple-output (MIMO) configurations [15].

The implementation of parallelism coupled with communication channels carries profound implications for the enhancement of multi-channel efficiency, as explained by the Shannon–Hartley theorem [16]:

[1]Department of Mechanical Engineering, National University of Singapore, Singapore 117575, Singapore.
[*]Corresponding author, e-mail: mpezgy@nus.edu.sg

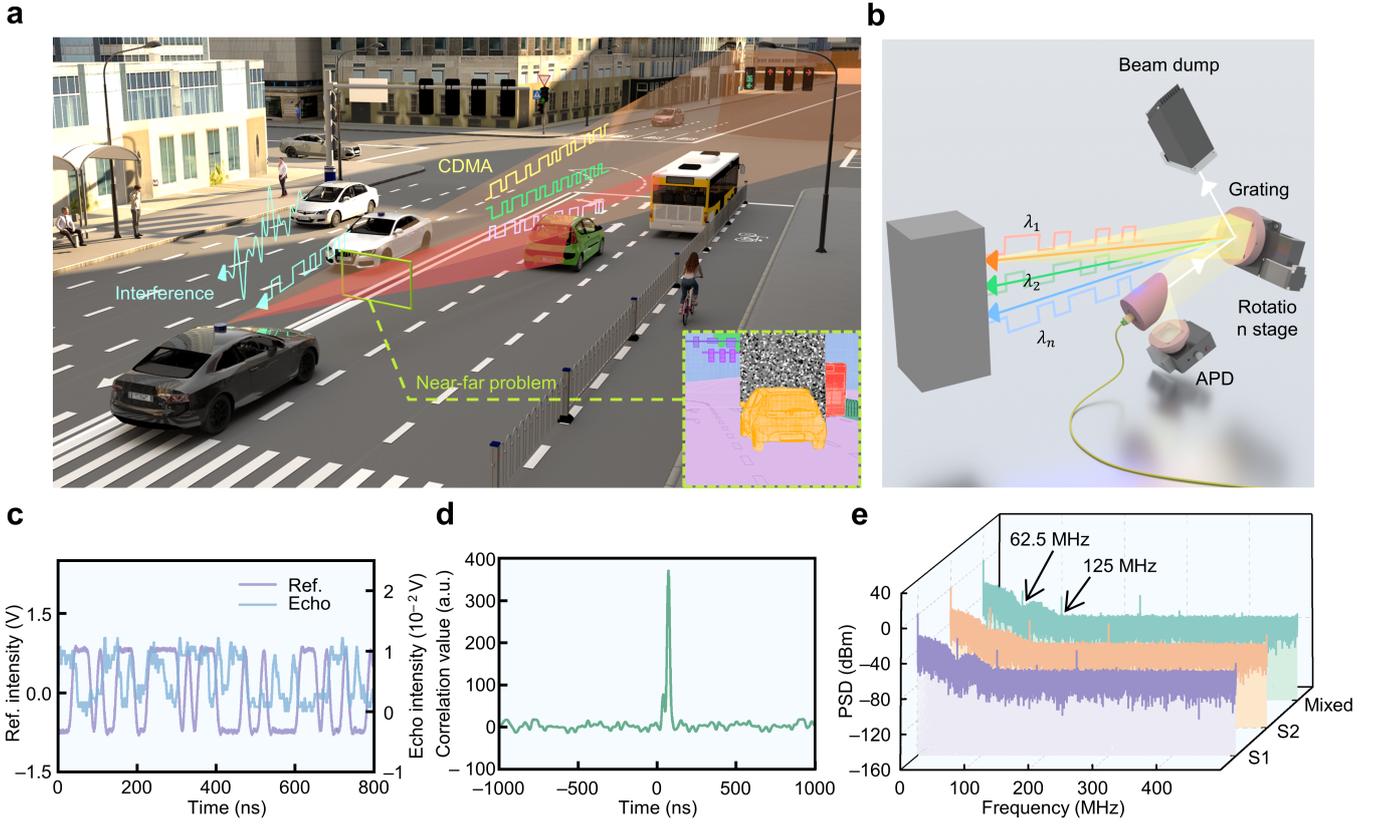

**Fig.1 | An identity-enabled parallel ranging CDMA LiDAR.**
**a,** Example of traffic conditions that exhibit multi-user interference and near-far problem (inset shows image confusion in segmentation caused by near-far problem). **b,** Coaxial architecture with wavelength multiplexing to enable simultaneous vertical emission and channel differentiation via CDMA. **c,** Partial plot of experimentally captured reference and echo signals. **d,** Partial correlation plot of the two signals received in **c**; peak time delay is used to determine distance information at scan point. **e,** Demonstration of spectral stability in spread spectrum modulation across parallel channels (mixed is composed of sequence 1 and sequence 2).

$$C = B \cdot log_2(1 + \frac{P_s}{BN_0}) \qquad (1)$$

where $C$ is the user capacity, $B$ is the bandwidth, $P_s$ is the average signal power, and $N_0$ is the additive noise intensity. By elevating the bandwidth or harnessing the spread spectrum, it is possible to increase capacity while adhering to the safety threshold for laser intensity [17]. Consequently, random modulated continuous wave (RMCW) LiDAR emerges as a notable alternative. This approach involves the modulation of carrier waveform with a stochastic amplitude pattern to enable distance determination by computing the time delays derived from the correlation indices between the echo and reference signals [18]. The adoption of pseudorandom sequence spread spectrum or broadband noise modulation, combined with coherent averaging, endow RMCW LiDAR with exceptional robustness against interference and speckle noise, facilitating a pivotal shift from single-channel applications [19-23] to multi-channel operations [7, 13, 24]. Nonetheless, in complex traffic scenarios, the challenge of selecting non-repetitive codes for a substantial number of users demands deeper consideration. Moreover, the near-far dilemma [25], originating from disparate target reflectance and signal attenuation with distance, is an issue that needs to be resolved in correlation-based ranging methods (refer to **Fig.1a**).

In this study, we present an optical code division multiple access (OCDMA) multi-channel LiDAR. It deviates from existing two-dimensional optical modulation systems with complex hard-limiting decoders [26] methods involving full-spectrum temporal pulse encoding [27]. Capitalizing on the unique correlation attributes and versatility of Gold sequences, our OCDMA design enables the easy implementation of direct-sequence spread spectrum (DSSS), and highlights a practical yet elegantly streamlined solution for comprehensive parallelization in detection and ranging. **Fig.1b** illustrates our proposed system where laser beams of various wavelengths, each modulated with a unique CDMA code, are combined into an optical fiber and directed to a diffraction grating via a fiber collimator. The first-order diffraction from the grating illuminates different heights of the scene with different wavelengths. All echo signals then retrace their original path, diffracted again by the same grating, and are received by a parabolic mirror and focused to an avalanche photodiode (APD). Simultaneous ranging in the vertical dimension is achieved statically by implementing a massively parallel wavelength-multiplexed CDMA scheme to correlate the mixed echo signal received from the APD with each reference CDMA code to decode the delays at different wavelengths. A 3D point cloud of the scene is obtained through a horizontal 1D push broom scan of the laser beams by rotation of the diffraction grating. Unlike systems that rely on true-random signals [7, 24], the engineered predetermined pseudo-noise (PN) binary codes in our approach obviates the need for high-speed photodetector arrays to capture reference signals for correlation and thereby reduces hardware cost and requirements. The integration of a diffraction grating scanner into our system provides multi-wavelength spatial multiplexing to mitigate peer-channel interference. It also simplifies a 2D rapid point scan to a slower 1D push broom scan, significantly reducing the inertia and dynamic deformation challenges in scanning mirrors [28, 29]. This innovation facilitates future miniaturization applications associated with microelectromechanical systems (MEMS) or chip-scale optical phased arrays (OPAs) [30-32] to enhance overall system efficiency and compactness. Additionally, each wavelength-multiplexed emission and receiver channel follows a coaxial architecture design, allowing all massively parallel spatially distributed echoes to converge onto a single receiving spot. This design, combined with code orthogonality for channel differentiation, enables the use of a single APD for simultaneous reception of multichannel LiDAR signals. Furthermore, this design precisely narrows the instantaneous FOV of the receiver to match the emission spot which ensures coherent scan and detection at identical angles. The ambient noise and external interference can hence be effectively minimized to enhance the SNR (**Sup.1**). Here, we demonstrate a CDMA LiDAR capable of achieving massively parallel multi-channel ranging. With a CDMA chip rate of 62.5 MHz limited by our current lab equipment, the system attains a depth resolution of 25.68 cm and a theoretical angular resolution of 0.03°. Pixel confidence is determined through correlation analysis. A control matrix approach is implemented to eliminate the near-far problem. A CDMA multi-user identification code scheme is further implemented to overcome the issue of multiple LiDAR signal interferences. Experimental results demonstrate its proficiency in accurately mapping 3D point clouds at a distance of 10 m, highlighting its rapid and precise parallelized ranging capabilities. This advancement heralds the future miniaturization of such a massively parallel muti-channel LiDAR system using MEMS-based vibrational grating scanners [33].

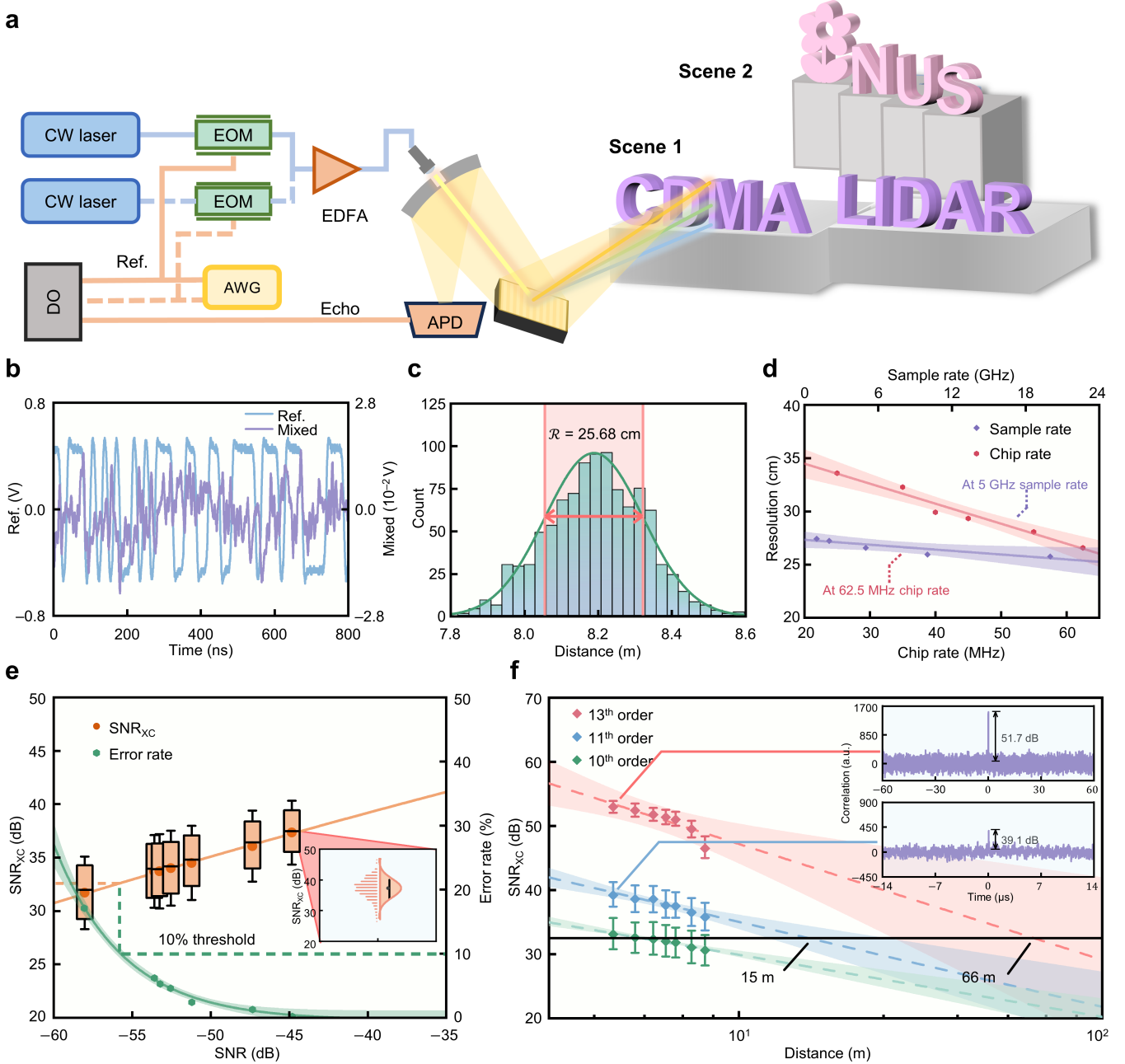

**Fig.2 | Ranging performance.**
**a,** Schematic of the experimental setup and system ranging test with targets at various distances for 3D imaging. **b,** Sequential signal superposition: raw data of mixed echoes from all 40 channels and reference signal from a single channel. **c,** Resolution analysis: histogram of repeated single-point detection and the depth resolution is defined as $\mathcal{R}$. **d,** Impact of chip rate and sample rate on resolution; enhanced resolution is observed with increased chip rates. According to our predicted trends, millimeter-level resolution can theoretically be achieved with a 200 MHz chip rate. **e,** $SNR_{XC}$ as a function of SNR. This plot is derived by adjusting the gain of the EDFA to produce various SNR levels. In our proposed massively parallel multi-channel detection scheme with a single-element detector, all other channel signals are treated as noise for a selected channel. With various SNR, ranging results are recorded and error rates are compared against known ground truths, which are also plotted in the figure (still workable in −50 dB testing environment). An $SNR_{XC}$ of approximately 32.5 dB, corresponding to a 10% error rate, serves as the threshold for data quality. The points indicate the median, the box spans quartiles two and three, and the whisker length corresponds to the standard deviation range. **f,** Predicted ranging limits for varying spread spectrum code lengths (95% confidence band). This figure demonstrates the estimated maximum ranging distances by varying the lengths of spread spectrum coding sequences and measuring the error rates at different target distances within a predefined range. All experiments are conducted with a constant output power. The insets highlight two contrasting $SNR_{XC}$.

**Gold sequences in parallel-ranging mechanism**
Rooted in RMCW fundamentals, this system modulates each channel independently using a uniquely generated Gold sequence from selected m-sequence pairings of two linear feedback shift registers (LFSRs). Gold sequences are well known for their minimal cross-correlation and sharp auto-correlation peaks [34, 35] (**Sup.2**). Their bipolar nature, represented by −1 and 1 chip values, simplifies modulation by eliminating complex amplitude or frequency adjustments [36]. Integration of field-programmable gate arrays (FPGA) alongside electro-optic modulator (EOM) can further enhance CDMA chip rates to the gigahertz range [37]. Strategic selection of preferred m-sequence pairs allows the generation of a vast number of quasi-orthogonal codes, facilitating massively parallel channel detection in mixed asynchronous signal environments. The correlation between the echo signal and its predetermined reference code is distinct (**Fig.1c,d**), enhancing signal clarity and system accuracy. The time delay of the correlation peak represents an estimate of the round-trip travel time, $\Delta\tau$, of the laser to the target, hence providing a distance measurement, $d$, given by:

$$d = \frac{c \times \Delta\tau}{2}, \ 0 < \Delta\tau < \tau_{max} \qquad (2)$$

where $c$ is the speed of light and $\tau_{max}$ is the symbol period that determines the maximum ranging distance without ambiguity. $\tau_{max}$ depends on the code length of the signal and the chip rate. The Gold sequence, employed as the spreading code, exhibits a power spectral density (PSD) similar to m-sequences, as depicted in **Fig.1d**. This sequence is characterized by a more pronounced central-frequency symmetric primary lobe and progressively diminishing side lobes, ensuring a more explicit spectral representation (**Sup.3**). Its broadband characteristic is contingent on the chip rate, which delineates the boundaries of the main lobe, allowing for a bandpass noise filter to be implemented. This approach of maintaining a non-expanded spectrum range, even with channel stacking, can accommodate a large user base within a finite bandwidth resource and optimize spectral efficiency.

**CDMA ranging performance**
The ranging performance for the experimental setup depicted in **Fig.2a** is thoroughly evaluated, and the detailed instrumentation and operational procedures are described in **Methods**. Confronted with instrumental limitations, a tunable laser is used to generate a series of coded laser channels to emulate the proposed parallel ranging process. This setup enables the sequential production of 40 distinct laser channels in the C-band and L-band to achieve an angular resolution of 0.03° in the height direction. An EOM modulates each laser channel with a corresponding orthogonal Gold sequence code from an arbitrary waveform generator (AWG). Signals carried by various wavelengths are transmitted through a polarization-maintaining (PM) fiber, amplified by an erbium-doped fiber amplifier (EDFA), and collimated by a fiber collimator with a 0.87 mm waist diameter into parallel beams. The parallel beams of different wavelengths are then diffracted by a ruled diffraction grating to different directions. The free space echo signals are captured by a parabolic mirror and an APD, converted to electrical signals, and simultaneously recorded with the AWG reference signal by a digital oscilloscope (DO). To demonstrate massively parallel ranging with 40 wavelength channels in the vertical direction, a specific post-processing approach is employed, which involves the aggregation of signals from all sequentially obtained wavelength channels to emulate the actual APD signal when all channels are transmitted and received simultaneously. The specific delay of each channel is then obtained by correlating the mixed APD signal with its reference code, as detailed in **Fig.2b.** It should be noted that because the system noises have been added 40 times in this post-processing approach, it produces a higher noise level than in actual parallel ranging where all channels are received simultaneously [27, 38, 39] (**Sup.4**). Additionally, the imperfections of a ruled grating may cause ghost lines [40]. However, another advantage of the proposed CDMA ranging scheme is that such extraneous correlation peaks can be accurately identified in the time delay after correlation with the reference code, allowing for the removal of any misjudgment from spurious peaks (**Sup.5**)

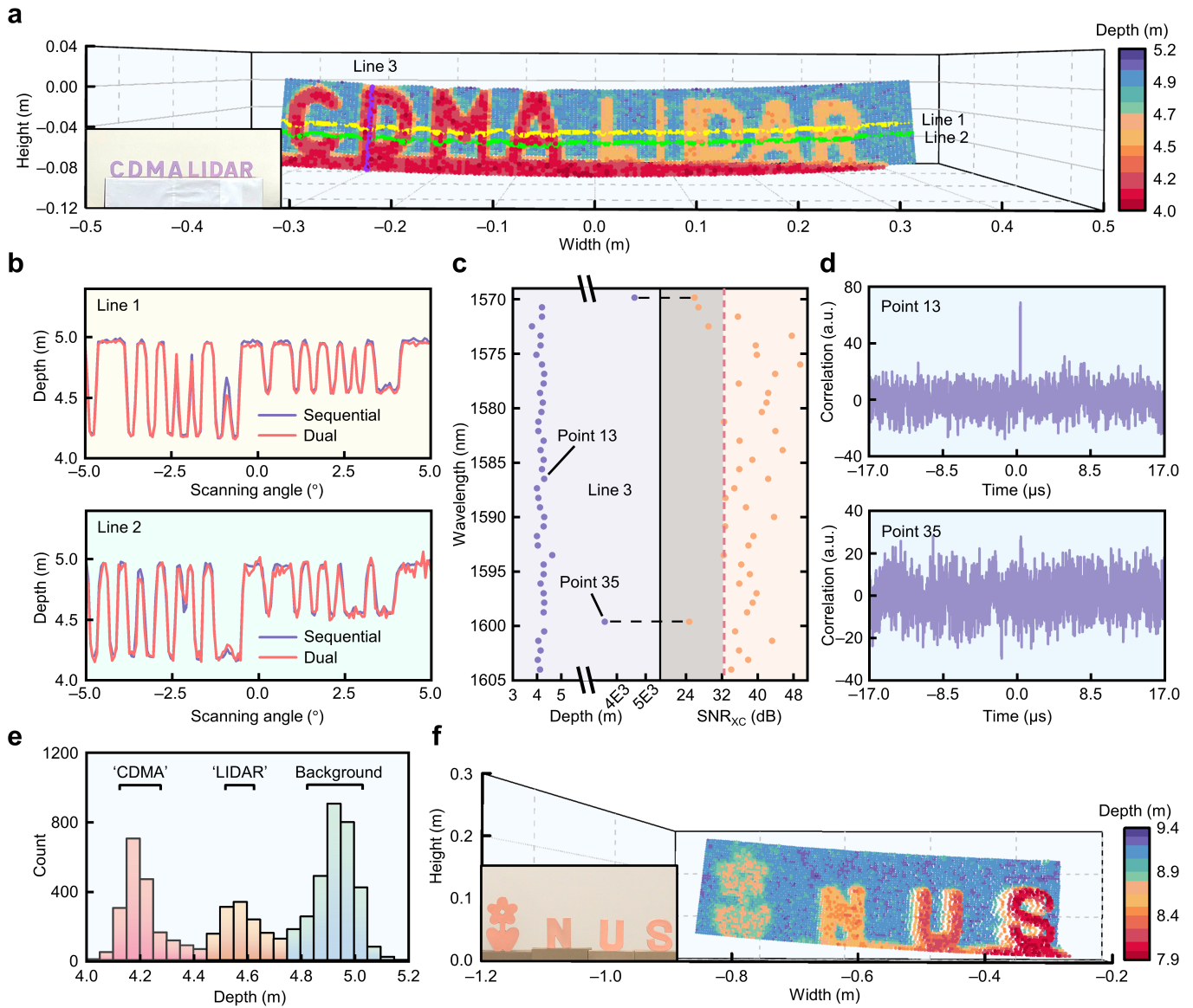

**Fig.3 | 3D imaging results.**
**a,** Reconstructed LiDAR 3D point cloud for the target at 5 m (a photo of the scene is shown in the inset) with an L-band EDFA. Yellow (line 1) and green (line 2) lines indicate two horizontal scans at identical wavelengths to compare the effects of simulated sequential mixing and actual simultaneous measurement. The purple line (line 3) shows 40 different channels at the same grating rotation angle to assess depth and image quality in parallel ranging. **b,** Comparison of sequential mixing and dual channel simultaneous ranging results at 1587.500 nm (line 1 in **a**) and 1591.875 nm (line 2 in **a**). **c,** Parallel ranging measurement of depth information from all channels in one column with their corresponding $SNR_{XC}$ (line 3 in **a**). False depth judgments with poor data quality at the edges of the object correspond to low $SNR_{XC}$. **d,** Specific $SNR_{XC}$ profiles for high-confidence point (point 13) with correct depth information and low-confidence point (point 35) with incorrect depth information. **e,** Histogram of scan frame shown in **a** with the three apparent primary clusters. **f,** The reconstructed 3D image for the target at 10 m (photo in the inset) with a C-band EDFA.

---

After calibration (**Sup.6**), the histogram of the distances for 1000 repeated measurements of a single channel ranging on the same test target is plotted in **Fig.2c**. The depth detection of the system, influenced by a delta-function-like correlation response [18] and other impact factors, predominantly matches a Gaussian distribution. Here, we propose using twice the precision to describe the system depth resolution. **Fig.2d** offers an estimate of the depth resolution when impacted by chip rate and sampling rate (**Sup.7**). While an increase in channel numbers may initially lead to signal degradation and affect the error rate, the optimal choice of code order and system parameters can minimize the impact of these limitations in actual practice (**Sup.8**).

Subsequently, it is determined that the free space path loss (FSPL) of the beam intensity resulted in an exponential decay in the echo signal. System dynamic range evaluation hinges on the cross-correlation signal-to-noise ratio ($SNR_{XC}$) associated with its correlation values (**Sup.9**). **Fig.2e** delineates a distinct relationship between SNR and $SNR_{XC}$, highlighting the effect of distance-related signal attenuation. An adaptive reference threshold with a 10% error rate is provided to quantify measurement performance. The attenuation threshold is also indicated in **Fig.2f** and predicts the maximum measurable range under prevailing lighting conditions. Targets are positioned at the measured intervals for this analysis, and **Fig.2f** also suggests that extending measurement time, akin to other coherence ranging, can expand this upper boundary [24], potentially achieving up to 66 m with a 13$^{th}$-order code.

### 3D imaging

To validate the performance of the prototype, two representative scenes are set up and tested, as shown in **Fig.2a**. The targets in scene 1 are the model letters "CDMA LIDAR" placed at approximately 5 m, and the targets in scene 2 are the model letters "NUS" and a model flower placed at approximately 10 m. 3D point clouds are captured with a 40 × 160 pixels resolution while maintaining the same vertical FOV angular span that is constrained by the wavelength range of our EDFA. The horizontal FOVs in scene 1 and scene 2 are −5° to 5°, and −5° to 0°, respectively. The 3D imaging result of scene 1 is shown in **Fig.3a**, where 40 wavelength channels with different Gold sequences are emitted and received sequentially and then mixed to emulate the actual multi-channel echoes on the single-element detector. The mixed echo is then correlated with each reference code to estimate the respective time delay and therefore the target distance of that channel. To mitigate the artificial excessive noises generated in the sequential mixing processing (See **Sup.9**), each mixed echo is recorded 10 times and averaged to ensure a stable and reliable data representation in the point clouds shown in the figure [41]. Considering the number of light sources and EOMs available, multi-channel simultaneous ranging is further demonstrated by two lasers emitted at 1587.500 nm and 1591.875 nm, each modulated with their unique code. They are combined and sent into a single fiber to emit and scan the same targets in scene 1, and their echoes are received simultaneously on the APD. Ranging results after cross-correlation are indicated by the yellow and green lines in **Fig.3a**, corresponding to the 1587.500 nm and 1591.875 nm results, respectively. **Fig.3b** further compares the ranging results for the two measurement approaches, with the purple line showing the results for sequential mixing of all channels and the red line showing simultaneous ranging results of two wavelengths. This shows that while superposition introduces more noise as previously mentioned, the results closely mirror actual conditions. Therefore, the subsequent experiments employ a single laser source to simulate the superposition of 40 channels, aiming for consistency with real-world outcomes. **Fig.3c** focuses on parallel detection within the same column (purple line in **Fig.3a**) and demonstrates the decoupling of information, individual channel depth measurement and its corresponding SNR$_{XC}$ from an all-channel mixed signal. This decoupling process also reveals how edge effect at object boundaries [42, 43] can lead to potential pixel confusion or coherence error. Further, **Fig.3d** shows the difference between a high-confidence pixel with a decent SNR$_{XC}$ value (clear and narrow cross-correlation peak), and a low-confidence edge point with low SNR$_{XC}$ value (cross-correlation curve is broad and noisy). A threshold in SNR$_{XC}$ is established to refine data accuracy by excluding erroneous points, which are then corrected by linear interpolating adjacent distance values. **Fig.3e** presents a concise histogram of the observed scene where the objects are clearly categorized into three distinct clusters based on their distance. The 3D imaging result of scene 2 is shown in **Fig.3f**. Space limitations in the lab reduced the available scanning field, creating a unilateral bow-like shape in the image. The uneven distribution of the scan points caused by the scanning grating is seen as distortion and is fixed by a unique algorithm to map distances precisely (**Sup.10**). After the distortion is corrected, high-quality 3D imaging is exhibited with distinct edges and depth variations, showcasing the stellar spatial depth resolution of the system even at a further distance.

### Identity-enabled ranging

The previous discussion highlighted the practicality of conducting parallel measurements with CDMA LiDAR systems, which is particularly significant in complex road traffic situations. Such scenarios may introduce challenges like suppression of interference from dense LiDAR signals and requirement for extensive high-confidence parallel measurement capability to accommodate a high volume of users. Hence, solutions to mitigate the increased risk of sequence code collision and signal redundancy are required. In response, we propose two unique solutions tailored to our system to address these challenges. The first strategy involves increasing the generated polynomial order to form a more extensive set of sequence combinations. For instance, elevating from the 11$^{th}$ to the 13$^{th}$-order nearly quadruples the orthogonal code candidates from 2049 to 8193, which reduces the risk of collision among different users selecting from the sequence pool. The second approach employs user identification (UID) code to capitalize on the inherent communicative properties of the unique spread spectrum codes of the system based on the nature of CDMA binary modulation. By leveraging modulation techniques such as polarity reversal (**Fig.4a**) or binary phase shift keying (BPSK), each symbol period is appended with additional bits carrying identity information to effectively enable a secondary encoding scheme for enhanced safety operation. Here, the identity sequences for secondary encoding are constructed using non-repeating binary codes. For example, three consecutive cycles of an 11$^{th}$-order Gold sequence can be used to incorporate three bits of identity information. This method yields four distinct identity codes (111, 000, 101, 100) for four users that cannot be derived from one another through phase shift, thus eliminating ambiguity. It should be highlighted that the time duration for an 11$^{th}$-order sequence carrying a 3-bit UID is shorter than a 13$^{th}$-order sequence with no UID, yet provides practically the same amount of sequence candidates. During channel decoupling and ranging, the received signal is correlated with the original sequence carrying the UID. As sequences from various users may collide (i.e. multiple users using the same sequence), multiple peaks exist in the cross-correlation function. However, the distribution of correlation peaks in the positive and negative axes defines the UID of each user, as depicted in **Fig.4b**. Hence, each user can only correctly

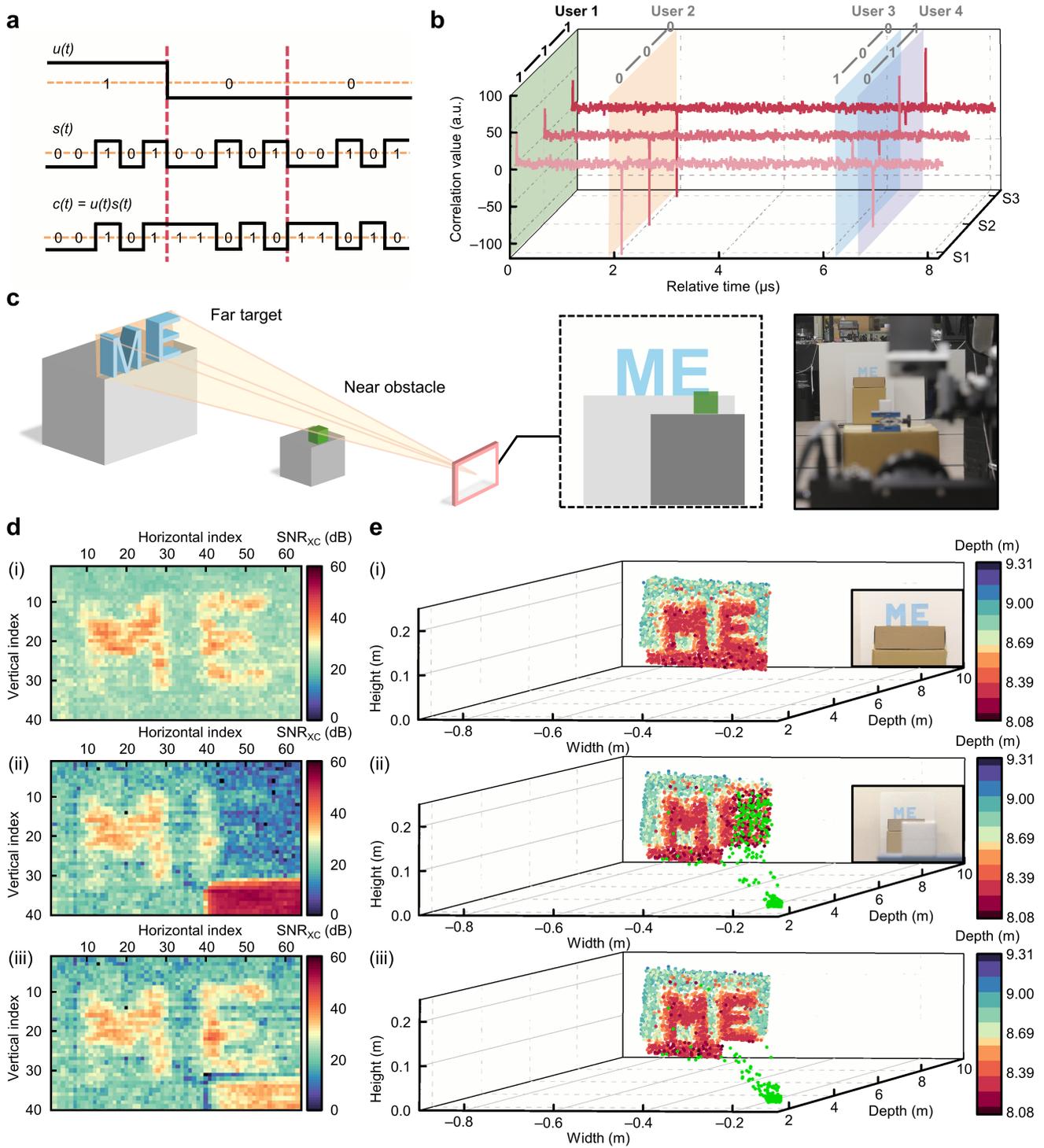

**Fig.4 | UID and solution to near-far problem.**

**a,** Schematic of spread spectrum code $s(t)$ carrying UID information $u(t)$. **b,** Cross-correlation from the perspective of User 1: Using unique UID codes to distinguish between four CDMA LiDAR users sharing the same Gold sequence (The full experimental signal is split into three consecutive symbol periods S1, S2 and S3). Users can only correctly compute its own distances with its local trigger signals. **c,** Schematic of experimental setup to emulate the near-far problem where a nearer block partially obstructs the scan of the farther model letters 'ME'. A photograph of the obstructed frame is on the right side. **d,** SNR$_{XC}$ heatmap showing pixel confidence and signal strength: **(i)** Unobstructed imaging reveals approximately uniform signal quality across the frame without near-far issues. **(ii)** Obstruction-induced near-far problem affects columns 40–64, with nearby object showing strong signal quality and distant pixels in the same column displaying low confidence. **(iii)** Image exhibits credible SNR$_{XC}$ values after laser intensity control is implemented, with minor edge effect on select object. **e,** Point cloud images: **(i)** Unobstructed model letters scan shows clear imaging without blur. **(ii)** Obstruction-induced near-far problem with the blurred model letter 'E' represented by the green points that indicate the nearer block. **(iii)** Corrected point cloud with distinct representation of both near and far objects.

---

identify its own correlation peak and delay when the symbol periods carrying the complete UID information are received. Obviously, under static conditions (UID does not dynamically change), the sequence candidates are effectively quadrupled, thereby significantly reducing the probability of random selection collisions compared to simply extending the sequence length (**Sup.11**). Moreover, implementing UID keeps acquisition time relatively low since utilizing higher order codes necessitate the complete acquisition of the entire sequence before range information can be ascertained. Compared with a 13th-order Gold sequence, a system with a 3-bit UID and an 11th-order Gold sequence can more rapidly obtain range data upon the reception of the first 11th-order sequence. While this method carries a shallow risk of code collision, such occurrences can be promptly detected and corrected upon the reception and verification of the subsequent UID.

**Near-far problem mitigation**
Near-far problem is a critical concern in multi-user communication technologies, mainly due to the variable strength of the returned channel signals where some can overshadow others. This issue is more difficult to address in asynchronous parallel detection systems and is exacerbated by the unpredictable and fluctuating nature of the distribution of the echo signals. The complexity originates not only from static intensity differences inherent in the system architecture, but also from extrinsic factors such as the distance of objects, their reflectivity, and noise from other users. In correlation-based parallel ranging systems with multiple channels, overpowering signals from nearby objects combined with weak signals from distant objects can degrade the quality of the ranging data, especially for far objects. However, most existing random modulated parallel detectors have yet to adequately address this challenge, compromising signal clarity and data integrity.

Our proposed CDMA LiDAR parallel ranging system has an inherent advantage to address this problem. In addition to the time delay of the cross-correlation peak used for depth measurement, its peak value can also be utilized to compute the $SNR_{XC}$ for that data point, thereby generating a $SNR_{XC}$ heatmap to identify and subsequently resolve the near-far problem. To demonstrate this, we create an experimental setup to showcase the proficiency of our CDMA LiDAR. As depicted in **Fig.4c**, a small block near the scanner serves as the partial obstruction while scanning the model letters 'ME' (with the lower margin of the letter 'E' obscured). This setup adequately replicates the signal processing challenges posed by objects with strong signal strength and weak signal strength that co-exist within the same column in a parallel ranging system. We again employ the $SNR_{XC}$ of each individual data point as the metric for assessing image signal quality and pixel confidence. Initial results in **Fig.4d(i)** and **e(i)** indicate that the processed point cloud can distinctly display the model letters 'ME' in the absence of highly reflective obstructions. However, the placement of a white reflective block near the scanner, while not completely obscuring the letter 'E', induces a near-far problem within the signals that are mixed and processed for the same column. This phenomenon impacts the processing of signals from channels with relatively lower intensity, as evidenced by the lower mean values in the $SNR_{XC}$ heatmap in the top right corner in **Fig.4d(ii)**. This leads to inaccuracies and misinterpretations in the point cloud shown in **Fig.4e(ii)**. To solve this problem, a $SNR_{XC}$ heatmap obtained from the previous frame is used to delineate regions for closed-loop feedback adjustment (**Sup.12**) to specifically target areas with excessively high or low mean values. By reducing the laser output in regions with elevated $SNR_{XC}$ while maintaining the power in other channels, this approach circumvents the need to increase overall power, and effectively prevents the system from surpassing safety power thresholds. Based on this feedback control matrix, subsequent scan frames such as one shown in **Fig.4d(iii)** produces a more uniform $SNR_{XC}$ distribution to facilitate signal demodulation across mixed channels. Objects at various distances are therefore distinctly visible in the point cloud image in **Fig.4e(iii)**.

**Discussion**
In conclusion, an identity-enabled multi-channel parallel CDMA LiDAR is demonstrated. Unlike other RMCW LiDAR systems that rely on true random sequences, our design employs pseudo random Gold sequences. This eliminates the need for a photodiode array to acquire the reference signals and simplifies the multiple independent receivers in MIMO systems to a single APD capable of receiving and segregating mixed signals. Furthermore, CDMA is less demanding on hardware precision, making higher modulation rates more easily achievable. Building on the advantages of CDMA, UID-enabled ranging is further proposed to prevent multi-user interference, thereby enhancing safe operation for practical complex road traffic conditions. Near-far problem is effectively addressed by utilizing $SNR_{XC}$ heatmap obtained in a previous frame to regulate the output power of laser channels in the next frame. A dynamic feedback loop is thus created to regulate echo signals with huge disparity in strength. By using a diffraction grating, our prototype adopts a coaxial architecture that minimizes external interference. It also greatly complements the implementation of parallel CDMA channels in our wavelength-multiplexed system, providing 40 channels for static simultaneous scan in the vertical direction while rotation of the diffraction grating enables horizontal push-broom scanning. Our experimental results demonstrate impressive parallel ranging performances at 5 m and 10 m, revealing a robust and high-performance LiDAR design with enhanced safety and great adaptability to complex measurement environments.

Recent research highlight on-chip integration as another evolutionary direction of LiDAR [32, 44, 45]. The innovations showcased in this work and the advancement in integrated and silicon photonics are precursors to the integration of CDMA LiDAR into an on-chip, multi-channel parallel ranging system. Controllable laser arrays can be formed using narrow-band single-mode laser diodes [46] or tunable laser diodes [47]. Additionally, highly integrated multi-micro-cavity structures enable the generation of multi-wavelength or directly modulated lasers [48, 49]. A single laser source can also be used to produce an optical frequency comb to achieve on-chip multi-wavelength signal modulation [50], and further supported by high-speed modulators [51, 52] to fulfill system functions. Similarly, our use of a diffraction grating for wavelength-multiplexed ranging and the direction of light aligns seamlessly with OPA technology [30, 53-55], indicating a promising avenue to integrate inertia-free scanning in the miniaturization process. Given that the future trajectory of LiDAR complements our proposed technologies, the proposed identity-enabled CDMA LiDAR concept has great potential to be commercialized for a next-generation parallel ranging system that promises economic benefits and superior performance.

## Methods

### Experimental set-up

A tunable laser source (Santec TSL-510) with ultra-narrow linewidth is employed to output different wavelengths. A lithium niobate EOM (iXblue MX-LN-0.1) is coupled to an AWG (Agilent 33522A), which is pre-programmed with an 11$^{th}$-order Gold sequence spanning 2047 bits, to modulate the tunable laser at 62.5 MHz. A C-band EDFA (Thorlabs EDFA100P) amplifies the laser power up to around 100 mW to scan the target at 10 m (**Fig.3f**), while an L-band EDFA (Photonik EDFA) is employed for the target at 5 m (**Fig.3a**) to demonstrate a more comprehensive scan range. The amplified laser is output by a fiber collimator (Thorlabs F230APC-1550). A better optimized fiber collimator can improve the beam divergence angle and mitigates the effects of speckle blurring. It can also enhance the spatial resolution and reduces the required output power. The modulated laser beam output from the fiber collimator is diffracted off a plane ruled diffraction grating (Newport 33010FL01-550R) with 600 grooves per millimeter and a nominal blaze angle of 28.7°. The diffracted beams enable static multiplexed vertical scan, and horizontal scan is achieved by rotating the diffraction grating on a motorized rotation stage (Zaber X-RSW60A-E03). After scattering, all channel signals return on the same path and are converged by an off-axis parabolic mirror (Thorlabs MPD249V-M01), a meniscus lens, cylinder lens, and an aspheric condenser lens onto an InGaAs APD (Thorlabs APD430C).

We employ a dual-path parallel transmission for the experiment in **Fig.3b**. Another tunable laser source (Santec TSL-710) is introduced and an additional EOM is included for modulation. The two modulated signals are combined using a 50:50 fiber coupler (Thorlabs PW1550R5A1) and amplified with the previously mentioned L-band EDFA before output. A wavelength combiner can be used in future practice to ensure minimal mixing loss.

We conduct the characteristic tests multiple times as described. All 3D imaging results are averaged over ten repetitions to minimize uncertain jitter and mitigate systematic errors. For targets in the 5 m scan, all model sizes are within the scan height and no more than 6 cm in width. They are made of white polystyrene foam and spray-painted pale purple to approximate real-world reflectivity rather than special highly reflective materials (**Fig.3a** inset). For targets in the 10 m scan, the model letters are 10 cm in height and no more than 7 cm in width, while the model flower is 20 cm in height. They are made from cardboard (**Fig.3f** inset) and spray-painted pink.

To emulate various UID scenarios, our system sequentially captures signals from different users, which are then superimposed for analysis. We used distinct scenarios that more commonly scenario involves users at different distances detecting a target and concurrently receiving interference from signals of other users positioned at varying ranges (**Fig.4b**). The system demonstrated the exceptional capacity of UID to differentiate signals.

To illustrate the near-far problem, the nearer block in the experimental setup (**Fig.4d**) is made of white polystyrene foam and remains white for high-reflectivity, while the further model letters 'ME' are spray-painted pale blue for more ordinary absorption characteristics. The white block obscures just part of the model letter 'E' to showcase the problems that occur when processing the same column.

### Data acquisition and post-processing

In our study, we utilize a DO (Agilent DSO90404A) to achieve synchronized signal acquisition from both the AWG and the APD. Given the periodic nature of Gold sequences, the main concern is not the start point of the modulated signals but to ensure that the recording spans a complete code-length cycle. We automate all devices using LabVIEW, employing a unidirectional, non-repetitive scanning approach. Initiated from the top-left of the target, each wavelength from the tunable laser is modulated with the corresponding code. The DO acquires ten distinct and non-overlapping measurements in the near-instantaneous dwell time. After each column scan is completed, the diffraction grating is rotated by the rotation stage to scan subsequent columns until the entire area is covered.

In post-processing, echo data with clock information is overlaid using MATLAB. A bandpass filter is applied to the second sidelobe (125 MHz) to purify the signal in the spectral domain. Circular correlation is used to compute the correlation between the reference and mixed echo signals, which proves to be more suitable than linear correlation. This ensures that the peak signal is not misaligned and superimposed due to excessive physical distance, which leads to the peak value falling below the noise floor. While employing circular correlation introduces a theoretically bounded ambiguity period in range measurements, the significant advantage lies in applying the fast Fourier transform (FFT) for convolution in the frequency domain, which substantially accelerates data processing speeds. Indeed, this enhances processing speed and effectively reduces the algorithm time complexity [56]. Lastly, each depth information is corrected for distortion, producing the final 3D image.

# Contents

## Supplementary Note 1 | Coaxial architecture

Present scan-based light detection and ranging (LiDAR) systems can be classified as coaxial or non-coaxial architectures according to their optics. As illustrated in Fig. 1a, microelectromechanical systems (MEMS) scanning mirror-based and optical phased array-based (OPA-based) LiDAR sensors currently utilize a non-coaxial architecture with decoupled emitter and receiver optics. The scanning mirror scans the emitted laser pulses across the scene, and at the same time, the receiver uses separate optics with a large receiving aperture, and its field-of-view (FOV) covers the entire scene. In this way, the weak LiDAR return signal from a long-range can be enhanced by the large solid angle of the receiver optics. However, non-coaxial architecture has inherent disadvantages. Since the receiver's FOV has to cover the entire scene where the pulsed laser is scanning, this, unfortunately, also drastically increases the background noise level due to the ambient light reflected from the area other than the laser-illuminated spot. In addition, large FOV also requires large-sized photodetectors, resulting in increased dark current noise and reduced detector speed, both of which are undesirable for high-performance LiDAR sensors. At the moment, major LiDAR developers use an array of receivers in combination with complex electronics to overcome the problem of large detector size. Still, this inevitability increases the cost of the sensor.

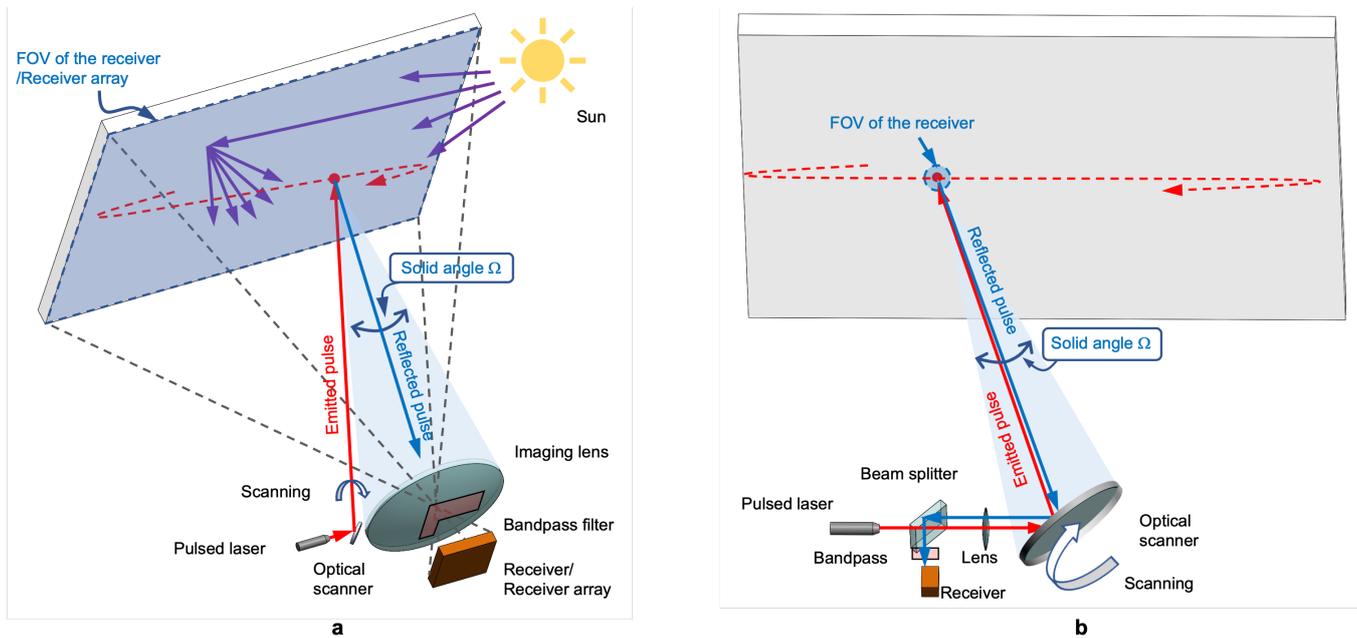

Fig. 1. Principles of (a) non-coaxial and (b) coaxial architecture.

In a coaxial architecture, the LiDAR sensor, emitter, and receiver are coupled to the same scanning optics (Fig. 1b). The FOV seen by the receiver is very small and is just enough to cover the spot illuminated by the emitted laser pulse. The small FOV of the receiver scans synchronously with the scan trajectory of the laser pulse. The coaxial architecture ensures that the receiver captures the light only at precisely the same angle as the outgoing light beam. In this way, ambient noise (sunlight) and signals from other LiDAR sensors are filtered out, resulting in a system with high signal-to-noise ratio (SNR). This is a preferred LiDAR design that presents an elegant technical solution to address the challenges of 3D perception in autonomous vehicles, which include limited laser power for eye-safe operation and thus low photon budget from long-range reflection from noncooperative targets, LiDAR signal drowned out by bright ambient light, as well as interferences from other LiDAR sensors present in the scene. Coaxial architecture is currently possible with a mechanical spinning LiDAR design, such as those commercially available from Velodyne. This work adopts a large aperture laser beam steering device with coaxial architecture. Hence, it has better ambient noise and multi-user interference suppression capability.

## Supplementary Note 2 | Generation and characterization of Gold sequence

The core principle of spread spectrum communications is the trade-off between frequency bands and SNR. This enhances the reliability of communication systems and allows multiple users to operate simultaneously within the same frequency band. The direct sequence spread spectrum (DSSS) technique uses a deterministic binary pseudo-noise (PN) sequence.

This sequence mimics the statistical properties of random noise but can be generated periodically and analyzed consistently [1, 2].

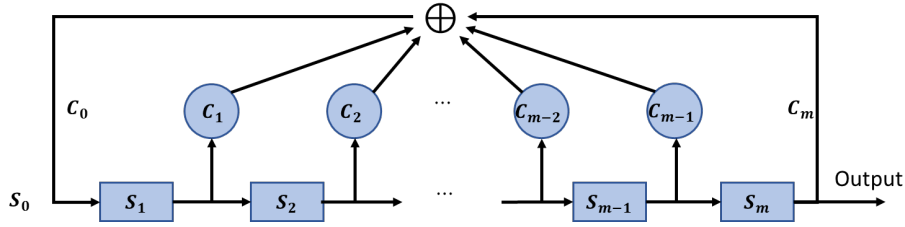

Fig. 2. LFSR for PN generator.

The maximum length pseudo-random binary sequence (PRBS) is generated by the linear feedback shift register (LFSR), as shown in Fig. 2, and is the m-sequence. By adjusting the switch binary state of each coefficient, different feedback polynomials $f(x)$ can be generated to produce a maximum sequence length $N = 2^m - 1$.

$$f(x) = C_0 + C_1 x + C_2 x^2 + \cdots + C_m x^m \tag{1}$$

Although m-sequences exhibit sharp autocorrelation and low cross-correlation characteristics, we further employ Gold sequences derived from m-sequences to obtain a larger number of orthogonal codes. To generate Gold sequences, two PN sequences, referred to as the preferred pair $u$ and $v$, are used. First, optimal generator polynomials are selected to create these sequences. By combining these sequences through operations such as XOR, one can derive Gold sequences that align with certain relationship-generating polynomials [3-5].

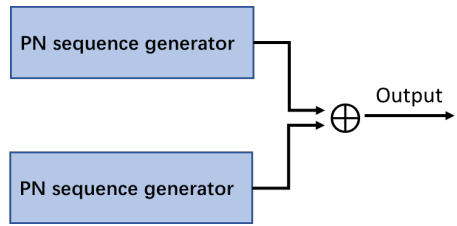

Fig. 3. Gold sequences generator.

As illustrated in Fig. 3, two confirmed pairs can generate Gold sequences (inclusive of the original PN sequence) by modifying the structural positional relationship between these two preferred pairs, applying vector shift operator $\Psi$, followed by a XOR operation.

$$G(u, v) = \{u, v, u \oplus v, u \oplus \Psi v, u \oplus \Psi^2 v, \ldots, u \oplus \Psi^{N-1} v\} \tag{2}$$

In our experiment, an 11[th] order Gold sequence was utilized, generated from one of the preferred pairs using the following polynomials,

$$f(x)_u = x^{11} + x^2 + 1 \tag{3}$$

$$f(x)_v = x^{11} + x^8 + x^5 + x^2 + 1 \tag{4}$$

In synchronization, the autocorrelation and cross-correlation functions of Gold sequences exhibit a triple-valued nature. This tri-valued correlation arises from the specific intercorrelation properties of any pair of deterministic m-sequences, as established by Gold and Kasami [6, 7]. Although Gold sequences do not always achieve the optimality set by Welch's lower bound [8], their favorable correlation properties, combined with the ability to generate a large set of diverse sequences, make them a popular choice in CDMA systems.

**Supplementary Note 3 | Cyclic correlation and power spectral density**
In spread-spectrum systems designed for multi-user operation, the behavior of correlation, encompassing both autocorrelation $R(\tau)_A$ and cross-correlation $R(\tau)_{AB}$, is of paramount significance [9]. Precisely, within ranging systems, the delay of a given signal is ascertained based on the peaks in cross-correlation, while autocorrelation is intrinsically tied to the system depth resolution. Autocorrelation and cross-correlation are described by:

$$R(\tau)_A \triangleq \int_{-\infty}^{\infty} A^*(t)A(t+\tau)dt \tag{5}$$

$$R(\tau)_{AB} \triangleq \int_{-\infty}^{\infty} A^*(t)B(t+\tau)dt \tag{6}$$

where $A, B$ are continuous functions, $*$ denotes the complex conjugation of function and $\tau$ is time delay. In our experiment, if $A, B$ are real and only exist in finite symbol period $T_0$, the cross-correlation is:

$$R(\tau)_{AB} \triangleq \int_{-T_0}^{T_0} A(t)B(t+\tau)dt \tag{7}$$

When the correlation calculation window slides beyond the symbol periods in the non-circular correlation of finite signals, the desired correlation peak decreases. This reduction occurs because the window extends into areas without signals, reducing the effective overlap and alignment between the signals (Fig. 4. linear correlation). Conversely, if we consider $A, B$ are both continuous periodic functions of symbol period $T_0$, the circular cross-correlation is:

$$R(\tau)_{AB} \triangleq \frac{1}{T_0}\int_{0}^{T_0} A(t)B(t+\tau)dt \tag{8}$$

Since the signal is periodic, the attenuation caused by the finite boundaries is avoided, so the correlation peak remains the same magnitude at any position (Fig. 4. circular correlation).

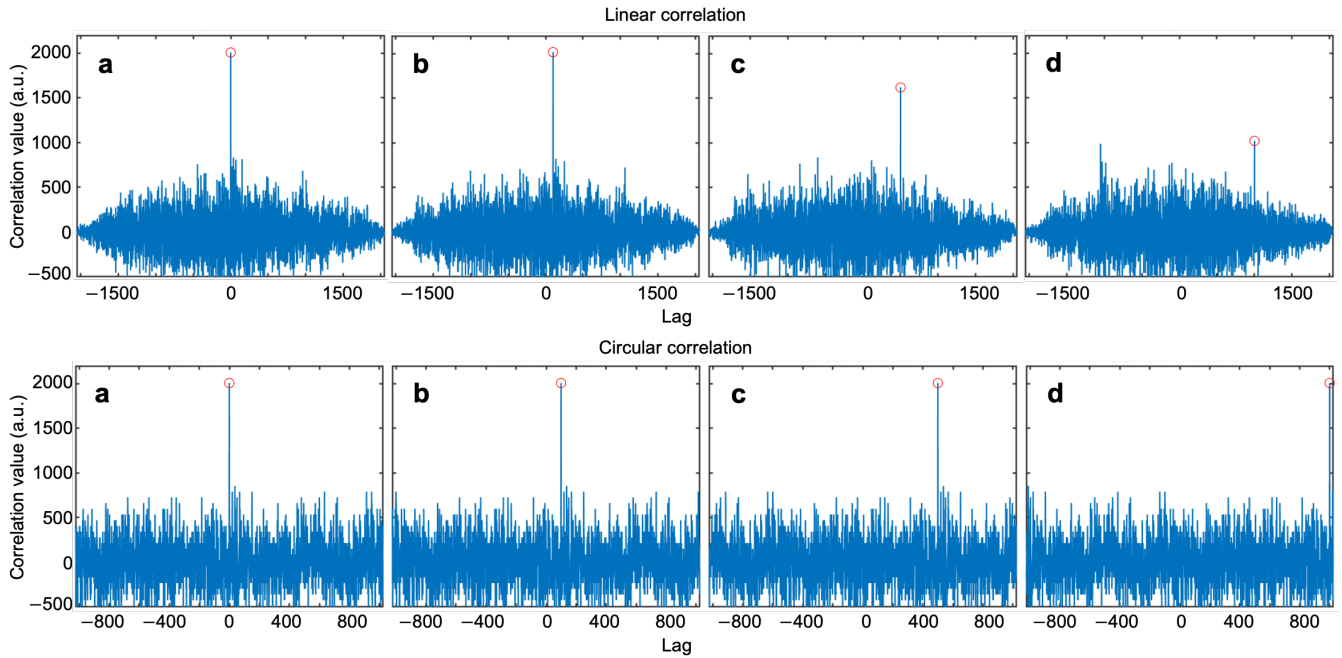

Fig. 4. Comparison between linear correlation and circular correlation, with lag occurring at (a) 0, (b) 100, (c) 500, and (d) 1000.

Simulated results are shown in Fig. 4, with four different lag conditions for linear and circular cross-correlation. Linear correlation extends the signal timeline to twice its original length, while in circular correlation, the negative semi-axis is a sequential continuation of the positive axis. The comparison shows that distinct peaks are evident at small lags in both scenarios. However, as the lag increases and approaches half the period length, the linear correlation is significantly influenced by finite boundaries, resulting in a rapid decay of the correlation peak. In actual asynchronous applications, this impact is more pronounced. Therefore, we implement circular correlation in our subsequent processing.

In practice, when we use a periodically repeating Gold sequence, the maximum detected unambiguous distance $d_{max}$ is expressed as:

$$d_{max} = \frac{c}{2}\underset{\tau}{\mathrm{argmax}}(\mathrm{xcorr}[s_r, s_e]) < \frac{c}{2}T_0 = \frac{c}{2}\frac{N}{\omega_c} \tag{9}$$

where, $s_r$ and $s_e$ are the reference and echo signals, respectively, $\omega_c$ is the modulated chip rate, and $N$ is the sequence length. The maximum unambiguous distance occurs at maximum time delay estimation of cross-correlation within one symbol period $T_0$. Moreover, by leveraging fast Fourier transform (FFT), correlation computations for long sequences can be translated into frequency domain operations [10], thereby increasing processing efficiency [11].

The depth resolution in theoretical analysis is typically determined by the full width at half maximum (FWHM) of the autocorrelation function for a singular code. Alternatively, when multiple measurements are taken at a constant point physically, the FWHM of distance distribution is employed [12, 13] to determine the experimental resolution. This convention stems from the delta function of the autocorrelation function (ACF). Specifically, for m-sequence, given that its autocorrelation demonstrates two values, it can be formulated as follows:

$$R(\tau)_m = \begin{cases} 1 - \frac{N+1}{T_0}|\tau - iT_0| & ,0 \leqslant |\tau - iT_0| \leqslant \frac{T_0}{N}, i = 0,1,2,\cdots \\ -1/N & ,otherwise \end{cases} \tag{10}$$

where the period $T_0$ is significantly large relative to an exceedingly small chip width $T_0/N$ (with faster chip rate). The correlation function resembles delta function $\delta(t)$. Consequently, at the zero-offset position, a minimal pedestal width is observed. Similarly, Gold sequences illustrate the same function-shape in zero-offset but three other values exist in other conditions [1].

$$R(\tau)_{Gold} = \begin{cases} 1 - \frac{N+1}{T_0}|\tau - iT_0| & ,0 \leqslant |\tau - iT_0| \leqslant \frac{T_0}{N}, i = 0,1,2,\cdots \\ -\frac{1}{N} \text{ or } -\frac{t(m)}{N} \text{ or } [t(m) - 2]/N & ,otherwise \end{cases} \tag{11}$$

$$t(m) = \begin{cases} 2^{(m+1)/2} + 1 & \text{odd } m \\ 2^{(m+2)/2} + 1 & \text{even } m \end{cases} \tag{12}$$

Those formulas suggest a straightforward program to increase resolution by simply increasing chip rate. Other factors affecting the distribution of this range include, but are not limited to, modulation overshoots, system jitter, and other interfering signals in asynchronous systems.

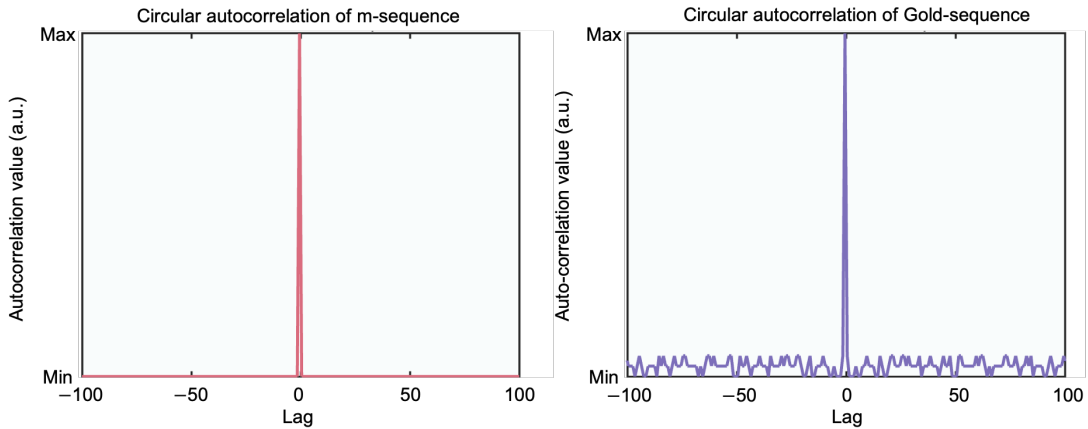

Fig. 5. ACF for m-sequences and Gold sequences (synchronous assumption).

According to Wiener–Khinchin theorem, the ACF $R(\tau)$ and the power spectral density (PSD) $P(\omega)$ of a signal $f(t)$ are a Fourier transform pair, it is possible to obtain the PSD of the Gold sequence by:

$$\begin{aligned} P(\omega) &= \int_{-\infty}^{+\infty} R(\tau)e^{-j\omega\tau}d\tau \\ R(\tau) &= \frac{1}{2\pi}\int_{-\infty}^{+\infty} P(\omega)e^{j\omega\tau}d\omega \end{aligned} \tag{13}$$

Since the bipolar spread spectrum code can be considered to utilize the same square wave modulation, the PSD is similar. Taking the m-sequence as an example, the theoretical power spectrum is approximately described as:

$$P(f) = \frac{1}{N^2}\delta(f) + \left(\frac{N+1}{N^2}\right)\sum_{\substack{i=-\infty \\ i\neq 0}}^{\infty} S_a^2\left(\frac{i}{N}\right)\delta\left(f - \frac{i}{NT_c}\right) \tag{14}$$

where $S_a$ is a sinc function related to the envelope profile and $T_c$ is the chip time. Based on theoretical formulas, we can simulate the ideal PSD distribution. In Fig. 6a, each specific sequence exhibits a symmetrical distribution, and the boundary between the main lobe and side lobes are determined by the chip rate, denoted as $\omega_c = 1/T_c$. As the chip rate increases, the spectral lines become sparser and the power is distributed over a wider bandwidth, thus more closely resembling noise characteristics.

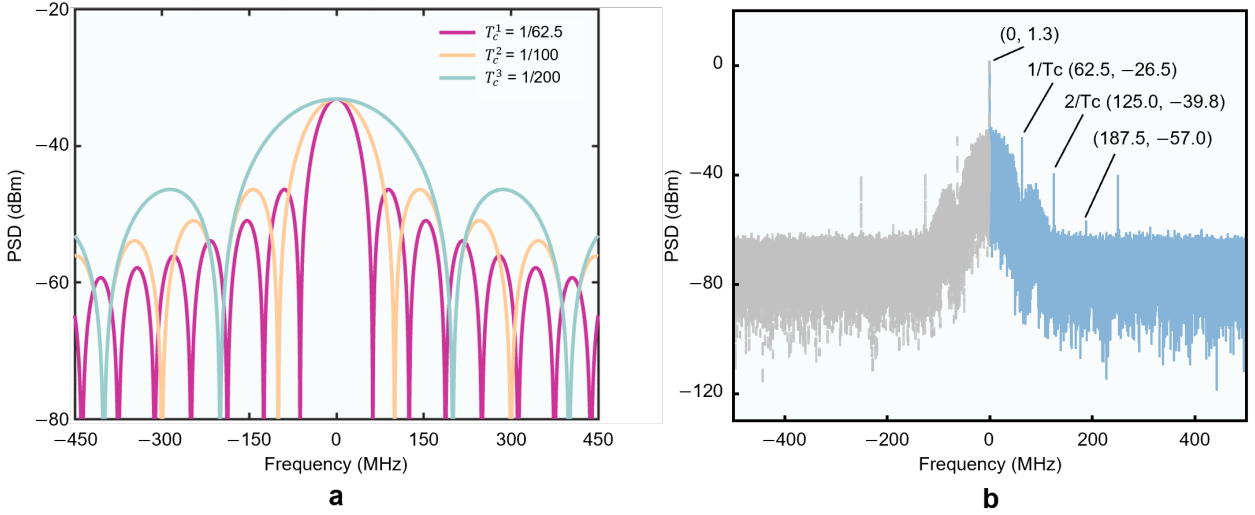

Fig. 6. Schematic of PSD: (a) Theoretical envelope of PSD for m-sequence; (b) experimental PSD for Gold sequence.

The Gold sequence exhibits a similar PSD distribution as shown in Fig. 6b. It can be observed that the boundary of the main lobe occurs at $1/T_c$, and then side lobes, followed by $2/T_c$. Due to limitations in our equipment, we can currently only achieve spread bandwidth with a modulation speed of 62.5 MHz. There is lobe boundary frequency leakage in the PSD of the experimental data caused by the sampling and modulation equipment, but the spreading effect is still clearly visible. Moreover, selecting the appropriate cutoff filter can effectively combat high-power noise from other non-target frequencies.

**Supplementary Note 4 | Assessing noise augmentation in multi-path superposition simulations**

Due to equipment constraints, we conducted sequential measurements by varying the laser wavelength. Subsequently, the raw echo signals are combined to emulate the scenario of concurrent measurements seen in real-world conditions. This approach poses no adverse effect when dealing with independent reference signals. However, the noise handling differs because post-processing involves the direct superposition of echo signals. The noise in the mixed signal, derived from sequential single-channel simulations, is significantly higher than in genuine parallel measurements. The primary sources of noise are erbium-doped fiber amplifier (EDFA), avalanche photodiode (APD), and environmental factors.

Our experiments employ specific wavelengths for EDFA amplification sequentially for different channels. However, in actual parallel conditions, multiple wavelengths within the target band are amplified simultaneously. Firstly, the amplified spontaneous emission (ASE) noise generated from a single low-power channel gain differs from that of near-saturated parallel power. The latter results in a better SNR under near-saturated or higher input power conditions [14]. Secondly, due to direct linear superposition, the wideband ASE noise generated across the entire gain spectrum is redundantly accounted for 40 times, whereas in reality, it should only be considered once. In addition to these factors, intrinsic thermal noise generated by the APD, dark current noise, and background light noise from environmental illumination can all be regarded as incoherent random noise sources. [15-17]. Now, by utilizing a simplified assumption to analysis, we suppose the echo received signal in the first sequential testing is $D_1$, the coherent net signal is $s_1$ and the noise is represented as $e_1$:

$$D_1 = s_1 + e_1 \tag{15}$$

Assuming the noise is incoherent and random with a zero mean, the Gold sequences, as pseudorandom signals, also exhibit this zero-mean property. Furthermore, the noise power is assumed to be approximately constant across each wavelength test, equating to its variance:

$$e_1 \sim \mathcal{N}(0, \delta^2) \tag{16}$$

Therefore, the actual 40-channel parallel measurement and noise power can be calculated as:

$$D = \sum_{i=1}^{40} s_i + e \tag{17}$$

$$E\{e^2\} = \delta^2 \tag{18}$$

In sequential superposition measurements, the post-mixed signal is:

$$D^* = \sum_{i=1}^{40} D_i = \sum_{i=1}^{40}(s_i + e_i) \tag{19}$$

$$E\{(e_1 + e_2 + \cdots + e_{40})^2\} = \sum_{i=1}^{40} E\{e_i^2\} = 40\delta^2 \tag{20}$$

A potential solution to this challenge is to leverage the properties of modulated signals. Though appearing random, the Gold sequences are deterministic codes produced by specific polynomials, enabling precise alignment in repetitions. For instance, consider the first measurement as an example:

$$D_1^{avg} = \frac{1}{40}\sum_{j=1}^{40} D_1^j = \frac{1}{40}\sum_{j=1}^{40}\left(s_1^j + e_1^j\right) \tag{21}$$

where $j$ refers to the repeated consecutive measurements of the same channel because $s$ is deterministic and $e$ is random and uncorrelated, hence:

$$D_1^{avg} = s_1 + \frac{1}{40}\sum_j e_1^j = s_1 + \tilde{e}_1 \tag{22}$$

$$E\{\tilde{e}_1^2\} = E\left\{\left(\frac{1}{40}\sum_j e_1^j\right)^2\right\} = \frac{\delta^2}{40} = \zeta^2 \tag{23}$$

where $\tilde{e}$ is the estimated average noise and power is represented as $\zeta^2$. The 40-channel mixed signal can be obtained as:

$$D^{**} = \sum_{i=1}^{40} D_i^{avg} = \sum_{i=1}^{40}(s_i + \tilde{e}_i) \tag{24}$$

$$E\{(\sum_{i=1}^{40} \tilde{e}_i)^2\} = 40\zeta^2 = \delta^2 \tag{25}$$

Consequently, by aligned signal averaging, the post-mixed signal can simulate the actual noise power level, eliminating the effect of repeated noise accumulation. However, we refrain from this processing in the experiments to simulate noise resilience and system robustness under heavy noise conditions. The results indicated that the performance is still satisfactory.

**Supplementary Note 5 | Analysis and mitigation of ghost lines**

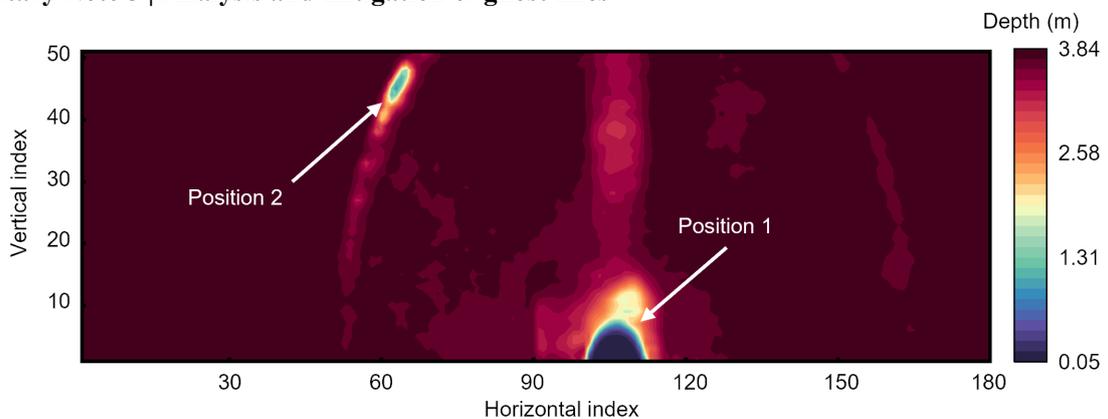

Fig. 7. Schematic of Lambertian reflection testing.

The presence of extraneous spectral lines caused by periodic irregularities in the spacing of grooves is commonly referred to as grating ghosts. Stray light in the context of a regular ruled grating is mainly attributed to inherent imperfections and manufacturing discrepancies that result in the diffraction of surface irregularities [18, 19]. This "ghost" phenomenon causes some of the diffracted light beams to deviate from the expected first-order diffraction direction. In practice, they are directly focused by the parabolic mirror and received by the APD. This type of stray beam, having the same sequence modulation as the emitted signal, appears as a coherent interference signal after post-processing.

A flat whiteboard is imaged at a fixed distance to conduct Lambertian reflection testing, as shown in Fig. 7. The effects of different errors are observed due to the presence of variable ghost interference under different wavelengths and grating rotation angles. The intensity of the grating ghost approaches its peak near position 1 and position 2 as indicated in the figure. Notably, while most pixels in the figure convey precise range information, this does not imply the absence of ghost interference in those areas. Variations in rotation angle and wavelength can generate spurious peaks lower than the actual echo signals. Although these spurious peaks may not overwhelm the correct signal, the phenomenon of ghost interference persists.

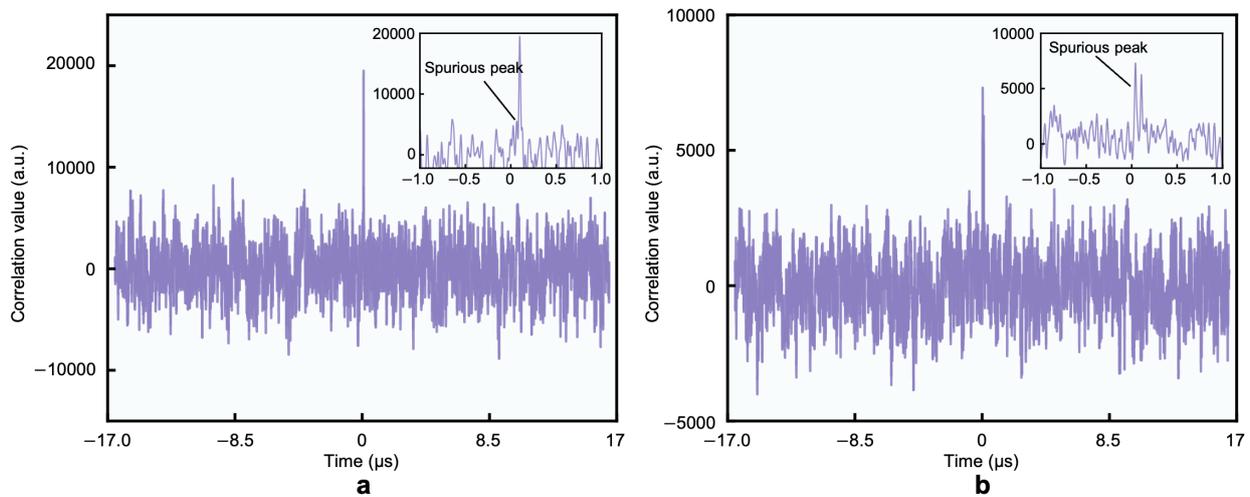

Fig. 8. Schematic of the spurious peak in correlation at (a) around 5 m and (b) around 10 m.

More specifically, as the ranging distance increases, the strength of the echo signal decays at a squared distance rate while the ghost intensity remains constant. Consequently, beyond a certain distance, the correlation peak from the ghost interference may surpass the actual peak, leading to ranging errors. Based on the 5 m scan results in Fig. 8, the distinctly powerful and discernible echo signal ensures that the delay index resides at the actual peak. However, at 10 m, the diminished echo signal intensity causes the correlation peak to fall below the spurious peak, overshadowing the actual correlation peak. To solve the grating ghost issue, holographic gratings can also be used to eliminate such artifacts [20, 21]. Alternatively, in our experiment, a more straightforward and intuitive countermeasure is to restrict the peak search range during correlation. Unlike the emitted beams, the location of the ghost interference does not change and can be considered a constant distance spurious object. Thanks to the discriminatory nature of CDMA, these spurious peaks can be labeled. When ranging far objects, the peak search range can be restricted after correlation to circumvent the ghosts. When the LiDAR measures nearby objects, the high intensity of the echo signal makes the range filter unnecessary, ensuring the proper retrieval of the peak index.

**Supplementary Note 6 | Calibration and signal alignment**
It is crucial to calibrate the system to achieve accurate ranging and imaging; otherwise, only relative distances among objects can be determined. However, pure theoretical calibration is challenging due to inherent fiber-optic delays and unquantifiable delays in instrument response. The utilization of a grating introduces a stationary zeroth-order beam. Typically, a beam trap is installed in the direction of this zeroth-order diffraction to eliminate its returning signal in standard operation (Fig. 9a). This feature can be harnessed during calibration to obtain an accurate target orientation. We rotated the grating 90 degrees from its ranging mode orientation, as shown in Fig. 9b, to ensure that the incident beam produces no diffraction orders apart from the zeroth. By covering the entrance of the beam trap with a flat object to reflect light, the measured distance from the grating origin to the trap is fixed to enable precise calibration.

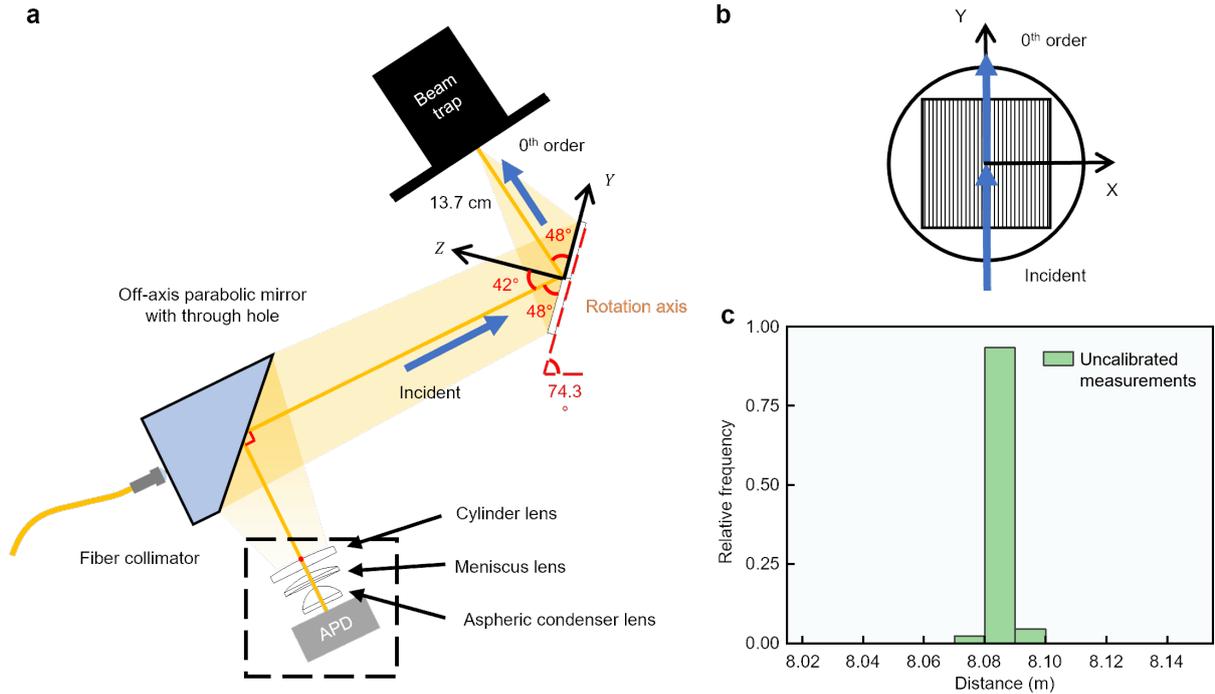

Fig. 9. Schematic of the calibration procedure: (a) experiment setup for calibration; (b) grating orientation; (c) histogram of one of the calibration (uncalibrated measurements).

Distinct system calibrations are conducted based on the different experimental configurations. One of the examples is presented in Fig. 9c. The distribution is derived by repeated measurements of the same target and normalization of the frequency of occurrence. Averaging the histogram data with distance distributions provide results that encompass system delay (uncalibrated measurements). Finally, subtracting the pre-determined distance of the fixed target yields calibration correction coefficients, which is utilized for absolute distance ranging. Since we use the $0^{th}$-order beam used for calibration does not change with wavelength, only one calibration is needed for each identical system setup.

**Supplementary Note 7 | Ranging resolution and precision**

Ranging resolution is a critical characteristic of LiDAR and defines the minimum depth separation in which a LiDAR system can distinguish two independent targets that are spatially close to each other. Mostly used in incoherent optical systems [22], radar [23] or traditional pulsed LiDAR systems that employ rectangular pulses, the return signal manifests approximately as a sinc function ($\sin(\pi x)/(\pi x)$). The theoretical distance resolution is achieved when the peak of one sinc function aligns with the first null (zero point) of the adjacent sinc function, ensuring no overlap ambiguity between the two targets. Much literature in RMCW LiDAR employ the traditional definition of depth resolution $\Delta R$ inferred from the FWHM of the ACF:

$$\Delta R = \frac{c}{2} \cdot FWHM = \frac{c \cdot T_c}{2} = \frac{c}{2B} \tag{26}$$

where $c$ is the speed of light, $T_c$ is chip time and $B$ is spread spectrum bandwidth. Under this definition, our theoretical traditional resolution is limited to 2.4 m due to the maximum rate of the function generator ($B = 62.5$ MHz). This obviously does not align with our experimental results (Main **Fig.3a, f**). Therefore, we propose a definition of depth resolution that is more applicable to our specific parallel RMCW LiDAR.

The traditional depth resolution $\Delta R$ represents a specific measurement scenario where the emitted light passes through a semi-transparent object and multiple reflections occur along the same optical path (e.g., two reflections separated by a distance $\Delta d$ in Fig. 10a). Applications that require harnessing multiple LiDAR returns along the same line-of-sight include environment monitoring and urban planning, where LiDAR sensors can penetrate vegetation layers and capture reflections from ground. In this case, signals with the same modulated sequence return with different time delays, producing multiple cross-correlation peaks in the same line-of-sight during post-processing. The critical point of traditional resolution definition is reached when the separation of adjacent peaks is equal to FWHM of the ACF (i.e. objects in the same line-of-sight can be separated only when $\Delta d \geq \Delta R$).

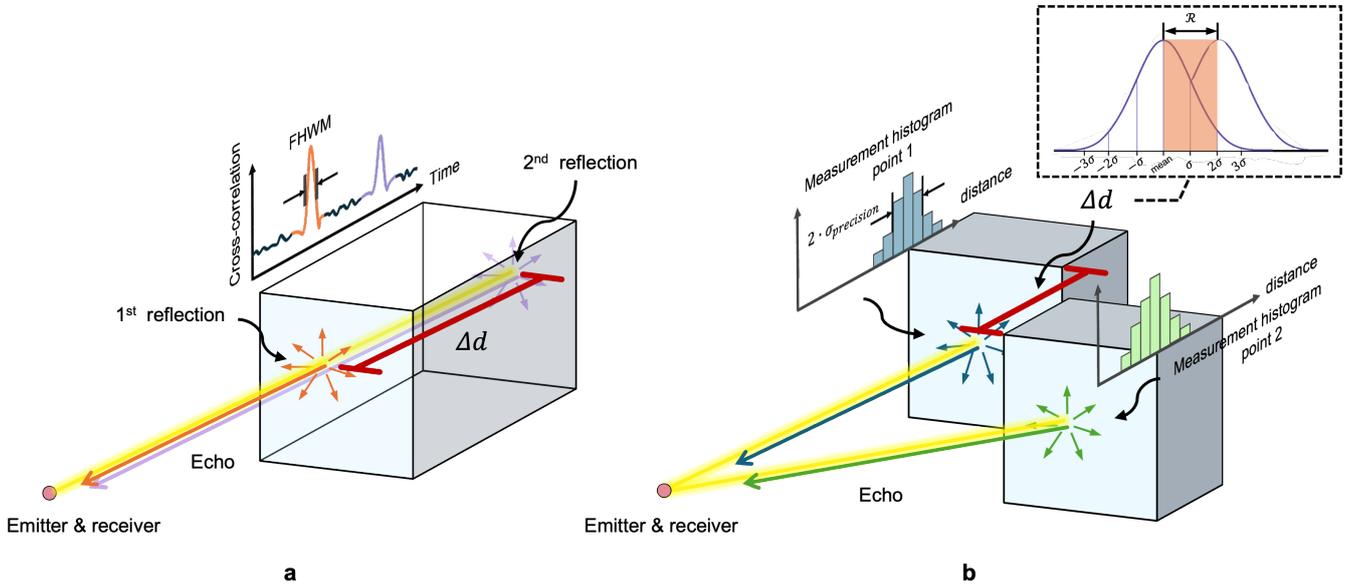

Fig. 10. Schematic describing (a) the definition of traditional resolution $\Delta R$ suitable for resolving multiple reflections along the same line-of-sight, and (b) our proposed precision-based resolution $\mathcal{R}$ suitable for resolving single reflections each from a different line-of-sight, with the inset describing the system depth resolution defined as twice the precision.

For many practical 3D imaging applications for autonomous vehicles and robots, resolving multiple reflections along the same line-of-sight as shown in Fig. 10a is not the primary goal of the system. Instead, resolving the distance of the first reflection from multiple different line-of-sights in the FOV of the LiDAR is considered more important for such applications, as shown in Fig. 10b. For a CDMA-based parallel LiDAR, each channel uses its own unique code. During modulation, the distance information of each channel is only valid when correlated with its local reference signal; otherwise, it appears as noise. Thus, different detection points can be considered within their respective line-of-sights, avoiding overlapped correlated peaks that would impair resolution between adjacent measurement points. Therefore, we propose a new resolution metric to better describe the novel system. As shown in Fig. 10b, the resolution $\mathcal{R}$ is still defined as the distinguishable distance $\Delta d$ between two adjacent reflected objects after sequential or parallel measurements. Here, we need to introduce the concept of precision. Mathematically, precision is the estimated value's standard deviation ($\sigma$) and reflects the impact of random noise on the estimation. Thus, repeated measurements of a single point produce a normal distribution range (as illustrated by the histograms in Fig. 10b), which indicates that the value of any single measurement fluctuates in this precision range. We propose that the depth resolution is represented by twice the precision ($2\sigma$). This ensures that the relative positional information of two adjacent points from different line-of-sights is not likely to be misinterpreted or blurred in most of measurements.

We follow an analytical formula introduced in Ref. [25-27] to estimate distance precision in RMCW LiDAR ranging:

$$\sigma_{precision} = \sqrt{\frac{6}{N_s}} \frac{1}{2\pi} \frac{\Delta R}{\sqrt{SNR}} \tag{27}$$

where $N_s$ denotes the number of sampled points in one symbol period, $\Delta R$ is traditional resolution defined in Eq (26), and $SNR$ denotes signal-to-noise-ratio (Supplementary Note 9). Thus, the proposed precision-based depth resolution can be expressed as:

$$\mathcal{R} = 2\sigma_{precision} = \sqrt{\frac{6}{N_s}}\frac{1}{\pi}\frac{\Delta R}{\sqrt{SNR}} = \sqrt{\frac{6}{N_s}}\frac{1}{\pi}\frac{c}{2B\sqrt{SNR}} \quad (28)$$

Corresponding to our experimental results, the resolution improves with a higher sample rate and SNR in a square root trend, as well as with an increase in chip rate. With our proposed definition, the actual performance can be more accurately estimated in system design and reduce unnecessary overestimation of system requirements.

**Supplementary Note 8 | Optimal management of signal attenuation and error rates**

According to the generation of Gold sequences described in a previous section, a preferred pair of polynomials can create $N+2$ orthogonal codes as mentioned before. Using a table of primitive polynomials (Table 1), multiple options exist to generate Gold sequences. However, sequences generated from different preferred pairs are not entirely orthogonal. Therefore, appropriate selection of primitive polynomials and pairs is crucial to avoid code conflicts and ensure good performance.

Table 1. Partial list of preferred pairs for the generation of Gold sequences.

| Degree | Preferred pairs of polynomials | Code candidates for each pair |
|---|---|---|
| 10 | [10,3,0] [10,8,3,2,0]* | 1025 |
|    | [10,7,1] [10,9,8,5,1] |  |
| 11 | [11,2] [11,8,5,2]* | 2049 |
|    | [11,9,1] [11,10,9,7,1] |  |
| 13 | [13,4,3,1] [13,10,9,7,5,4]* | 8193 |
|    | [13,10,9,7,5,4] [13,12,8,7,6,5] |  |
|    | [13,11,8,7,4,1] [13,11,10,5,4,3,2,1] |  |

* Represents the sequence generation polynomial used in our experiment and simulation.

In a practical free space LiDAR environment, increasing the number of channels per user and overlapping channels across different users can improve performance and enhance efficiency of the LiDAR. Although the Shannon-Hartley theorem relates user capacity (the total number of parallel measurement channels within the same spread spectrum at a given time) to signal quality and bandwidth, multi-user and multi-channel interference typically prevents reaching these theoretical limits. This study investigates how extending sequence lengths can improve $SNR_{XC}$ (refer to definition in Supplementary Note 9) and channel capacity. We assume all channels have equal power, but in fact channel interference from outside the FOV of interest would be of very low power due to the use of a coaxial structure.

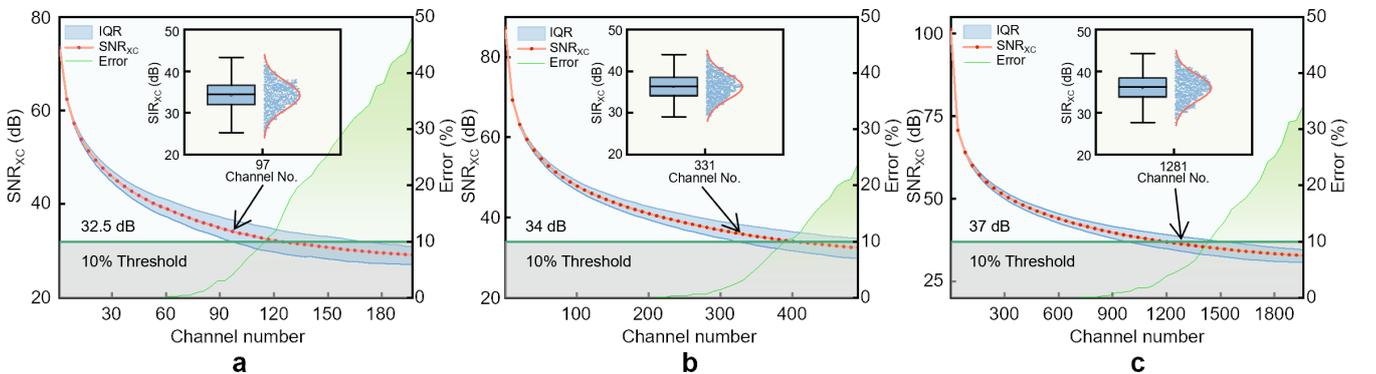

Fig. 11. Simulated channel capacity in different sequences of the (a) 11[th] order, (b) 13[th] order, and (c) 15[th] order.

To simulate the scenario of asynchronous CDMA reception, we sampled each chip period (one bit) in the form of 50 recording points. Considering that the distances to various detection targets can vary randomly, we used the Monte Carlo method to simulate each channel with 1000 random delays. The corresponding $SNR_{XC}$ trends and distribution patterns are obtained as the channels increased. On the other hand, to investigate the error rate, we determine the accuracy by comparing the calculated correlation peak index with the anticipated distance (ground truth). Notably, this error rate is used as a flexible threshold whose value can be adjusted based on the system and the desired accuracy in practice. For this simulation, an error rate of 10% is adopted as the preferred benchmark, similar to our experiment. The red sample dots in Fig. 11 represent simulation results with different number of signal channels. Concurrently, the blue shaded region shows the interquartile range (IQR) of the simulated data sets. Statistical analyses of data points are detailed in the insets. The visuals provide a clear distribution of the simulation results. When these results approximate a normal distribution, it supports the adequacy of the sample size used in the simulation. The cut-off $SNR_{XC}$ of approximately 32.5 dB for the 11$^{th}$-order sequence aligns with that measured in experiments as described in the main text. Additionally, it is observed that the $SNR_{XC}$ corresponding to a 10% error rate vary between different sequence orders. As the order increases, the maximum channel capacity improves. Specifically, the workable channel numbers for orders 11, 13, and 15 are roughly more than 125, 400, and 1200, respectively. Despite environmental noise in practical scenarios, the coaxial architecture of the system helps minimize the impact of multiple interference signals of equal intensity. An acceptable assumption in actual applications can be described with a modulation chip rate $\omega_c$ of 1 GHz, sequence length of the 13$^{th}$ order (i.e. $N = 2^{13}-1$), and 1000 spatial pixels in one image. Then, the ideal framerate for scanning can be calculated as:

$$f_{frame} = \frac{\omega_c}{N \times 1000} \approx 122 \text{ Hz} \tag{29}$$

Theoretically, current commercial technologies are fully capable of achieving real-time imaging in scenarios with a large number of users.

**Supplementary Note 9 | SNR, $SNR_{SS}$, $SNR_{XC}$ and multi-channel interference in an asynchronous system**

Various metrics are used to evaluate system performance and signal quality. In the raw mixed channel signal processing, the power ratio between the received echo signal of interest and the interfering signal of other channels (considered as noise) is similar to the universal definition, expressed as the signal-to-noise ratio (SNR). For our DSSS asynchronous system, considering the enhancement of spreading gain, the ratio of spread signal power to other interference and noise power is defined as spread spectrum signal-to-noise ratio ($SNR_{SS}$). In the subsequent cross-correlation, we introduce correlation-based peak to noise floor ratio as $SNR_{XC}$ to indicate the clarity and confidence of the correlation peak and data quality.

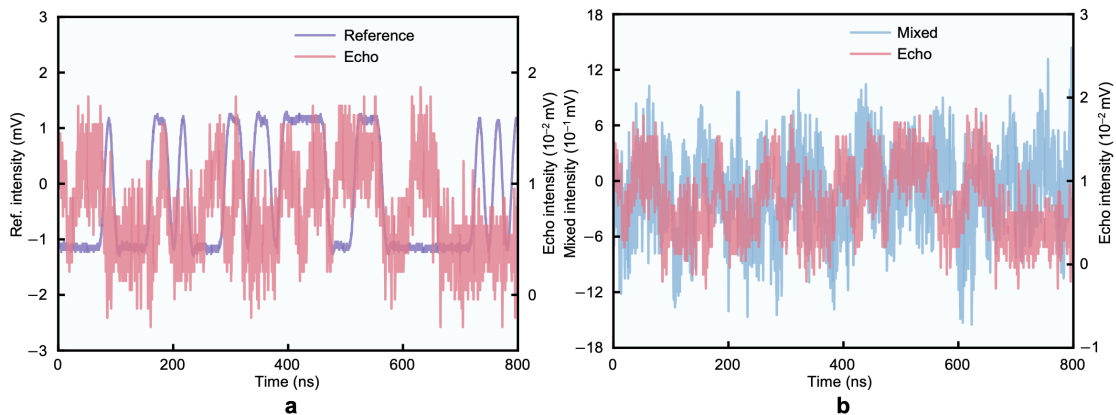

Fig. 12. (a) Delayed single-channel echo and its reference signal; (b) specific echo signal compared to all channels mixed signal.

For sequential channel-by-channel detection, one of the echo and its corresponding reference signals is displayed in Fig. 12a. As we demonstrate parallel ranging, an asynchronous mixed echo signal is shown in Fig. 12b. In the CDMA LiDAR, when mixed signals are received, return signals from channels other than the one of interest are considered interference noise. Due to the use of different pseudorandom sequences for all codes, they can be considered independent. Additionally, they exhibit the characteristics of uniformly distributed noise in the time domain. Therefore, we assume these signals have a zero mean, and their power can be represented by variance. Then, the raw mixed channel SNR can be expressed as:

$$SNR = \frac{P_s}{\sigma_\eta^2 + \Sigma_{i=1, i \neq s}^{K} P_i} \tag{30}$$

where $P_s$ is the power in the channel of interest, $P_i$ is the interfering channel power, and $\sigma_\eta^2$ is noise power (for example detector thermal noise), $K$ is the total number of channels. Consequently, SNR for parallel ranging drops to around –50 dB during experiments. This shows that the system is highly resistant to noise and interference and has the potential to operate in extremely low SNR environments compared to traditional incoherent LiDAR.

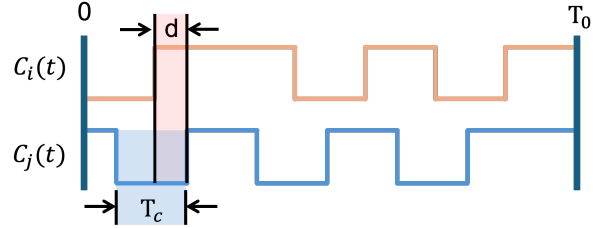

Fig. 13. Schematic of multi-channel interference in an asynchronous system.

In asynchronous CDMA systems, due to the inability to guarantee perfect alignment of sequences, the synchronization and code orthogonality performance is inherently degraded compared to synchronous scenarios. Such multi-user interference results in the characterization of A-CDMA as an interference-limited system. For any sequence of interest $c_i(t)$, the interference of another sequence $c_j(t)$ is illustrated in Fig. 13, where $T_c$ is chip period, $T_0$ is symbol period, and $d$ denotes shift delay. Therefore, the asynchronous cross-correlation is defined as:

$$r_{ij}^a(\tau) = \frac{1}{T_0} \int_0^{T_0} c_i(t) c_j(t + \tau) dt \tag{31}$$

Hence, the asynchronous cross-correlation can be expressed in terms of the summation of shifted synchronous cross-correlation:

$$r_{ij}^a(d) = (1 - f) \cdot r_{ij}(0) + f \cdot r_{ij}(-T_c) \tag{32}$$

$$f = \frac{d}{T_c} \tag{33}$$

The average value and power of asynchronous cross-correlation is:

$$E\{r_{ij}^a(d)\} = 0 \tag{34}$$

$$E\{[r_{ij}^a(d)]^2\} = E\left\{[(1-f) \cdot r_{ij}(0) + f \cdot r_{ij}(-T_c)]^2\right\} \tag{35}$$

$$E\{[r_{ij}^a(d)]^2\} = E\{(1-f)^2 \cdot r_{ij}^2(0)\} + E\{f^2 r_{ij}^2(-T_c)\} + E\{2f(1-f)r_{ij}(0)r_{ij}(-T_c)\} \tag{36}$$

The synchronous terms are similar to those described in Fig. 13, but without asynchronous gaps. By the property of PN sequences, the expected value for chips in correlation can be simplified as:

$$\begin{matrix} E\{c_i(p)c_i(q)\} \\ E\{c_j(p)c_j(q)\} \end{matrix} = \begin{cases} 1, & if\ p = q \\ 0, & if\ p \neq q \end{cases} \tag{37}$$

When $p = q$, the synchronous term is expressed as:

$$E\{r_{ij}^2(0)\} = E\left\{\frac{1}{N}\Sigma_p c_i(p)c_j(p) \times \frac{1}{N}\Sigma_q c_i(q)c_j(q)\right\} \tag{38}$$

$$= \frac{1}{N^2}\Sigma_p E\{c_i^2(p)\} E\{c_j^2(p)\} \tag{39}$$

$$= \frac{1}{N^2}\Sigma_p 1 = \frac{1}{N^2} \cdot N = \frac{1}{N} \tag{40}$$

For the first term on the right side of Eq (36), since $f$ is a random variable uniformly distributed between $[0, T_c]$:

$$E\{(1-f)^2 \cdot r_{ij}^2(0)\} = \frac{1}{3} \cdot \frac{1}{N} \quad (41)$$

Eq (36) can be finally expressed as:

$$E\{[r_{ij}^a(d)]^2\} = \frac{1}{3N} + \frac{1}{3N} + 0 = \frac{2}{3N} \quad (42)$$

where $N$ is the spreading gain that is approximately equal to the sequence length. Then, the SNR$_{SS}$ in an A-CDMA system is:

$$SNR_{SS} = \frac{N \cdot P_s}{\sigma_\eta^2 + \frac{2}{3}\sum_{i=1, i \neq s}^{K} P_i} \quad (43)$$

By introducing spreading gain, it can be seen that increasing sequence length enhances the SNR$_{SS}$ and further improves the detection range. In a practical system, echo signals from all channels are received compositely and simultaneously on a single receiver. It is impossible to estimate $P_i$ individually and therefore the SNR$_{SS}$ cannot be discriminately computed for each channel in the physical system. Thus, we propose to use the SNR$_{XC}$ in the experiment by quantifying the performance of the desired correlation peak.

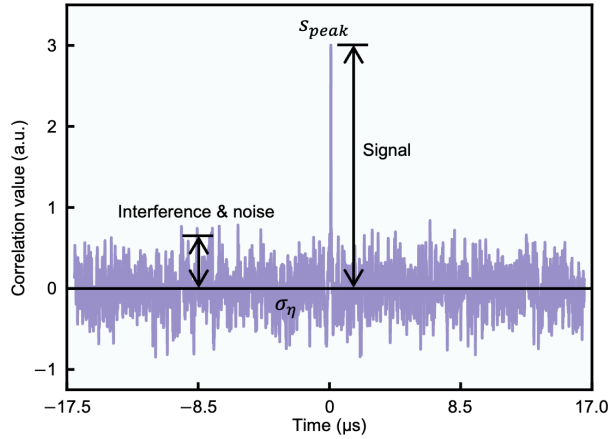

Fig. 14. Schematic of signal and noise in SNR$_{XC}$.

Fig. 14 shows the correlated result between the received mixed multi-channel signal and a reference signal from one of the selected channels. Aside from the distinct correlation peak representing the expected signal, all other channels manifest as noise (user interference and other noise). Thus, SNR$_{XC}$ can be defined as:

$$SNR_{XC} = \frac{s_{peak}^2}{\sigma_\eta^2} \quad (44)$$

where $s_{peak}^2$ is the power of the cross-correlation peak (i.e. signal power of interest), and $\sigma_\eta^2$ is the variance of the noise floor that includes both noises and interferences from other channels. The SNR$_{XC}$ is physically obtainable from the mixed channel signal from the receiver and enables us to determine ranging quality to address near-far problem, error filtering and maximum detection range prediction.

**Supplementary Note 10 | Grating operation principle and distortion correction in reconstruction**
For ease of understanding, we start with a typical in-plane rotation of diffraction grating as shown in Fig. 15a. The point of incidence on the grating serves as the origin of the grating coordinate system *XYZ*. The grating lines are parallel to the *X*-axis and span a spatial period dictated by $\Lambda$ in the *XY*-plane. We assume an incident collimated laser beam with wavelength $\lambda$ that illuminates the grating in the *YZ*-plane with incident angle $\theta_i$. The initial position with default rotation angle $\theta_r = 0$ is considered a standard origin and an anti-clockwise rotation around the axis is defined as positive. Due to diffraction, several diffracted beams with different orders (denoted by *m*) are produced, and the 1$^{st}$ order is evaluated in this case. The original scan point on the screen is shown as $P_0(x_0, y_0, z_0)$, and after rotating the grating clockwise by an angle $\theta_r$, the 1st order scan point is counterclockwise shifted (the zeroth order always remains stationary) to $P(x, y, z)$. In the experimental setup (Fig. 15c), the grating is mounted to the rotary stage at a predetermined angle, which takes into consideration the characteristics of the grating groove density, the input wavelength range, and elimination of the second-order diffraction beam. A simplified grating setup is shown in Fig. 15b, with the grating lines along the *X*-axis, and the *ZY*-plane represents the local coordinate system of the grating, with the origin at the center of the grating and the *Z*-axis perpendicular to the grating surface. In this figure, the *Z'Y'*-plane where the *Z'*-axis is perpendicular to the ground is the global coordinate system. Without grating rotation, the relationship between the incident angle $\theta_i$ and the 1st order diffraction angle $\theta_d$ is expressed as [28]:

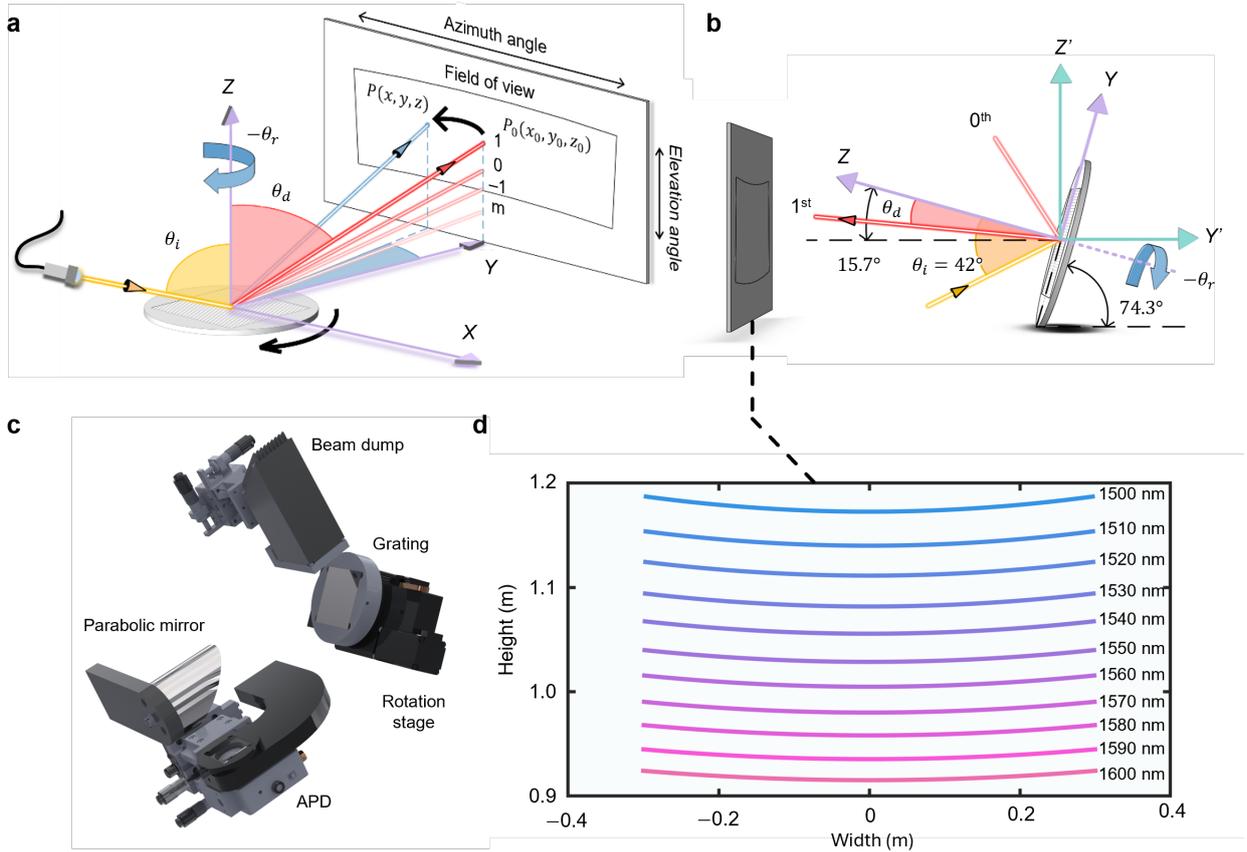

Fig. 15. (a) Schematic of a typical in-plane rotation of a diffraction grating for laser scanning; (b) grating orientation in experimental setup; (c) experimental setup; (d) simulation of scan distortion of first-order beam.

$$sin\theta_i + sin\theta_d = \frac{\lambda}{\Lambda} = G \qquad (45)$$

Our input wavelength $\lambda$ is limited to the C-band (1530 nm to 1565 nm) and L-band (1570 nm to 1605 nm), and the first order (*m* = 1) diffracted beam appears on the same side of the normal (*Z*-axis) as the incident beam.

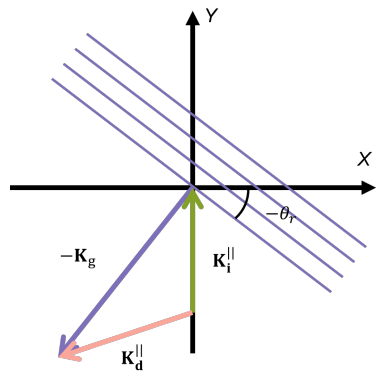

Fig. 16. Schematic of the projected wave vectors and the grating vector on the *XY*-plane.

When the diffraction grating is rotated by an angle $-\theta_r$ about its surface normal, the direction of the outgoing diffraction beam can be determined using the diagram shown in Fig. 16, where the wave vectors and grating vector fulfills the following relation:

$$\mathbf{K}_d^{\parallel} = \mathbf{K}_i^{\parallel} - \mathbf{K}_g \tag{46}$$

where $\mathbf{K}_i^{\parallel}$ and $\mathbf{K}_d^{\parallel}$ are projections of the incidence and diffraction wave vectors onto the grating plane, respectively. The grating vector denoted by $\mathbf{K}_g$, possesses a magnitude equal to $2\pi/\Lambda$, and the direction is perpendicular to the grating lines. Therefore, the components of the diffracted wave vector can be obtained as:

$$\begin{bmatrix} K_d^X \\ K_d^Y \\ K_d^Z \end{bmatrix} = \begin{bmatrix} -\frac{2\pi}{\Lambda}\sin\theta_r \\ \frac{2\pi}{\lambda}\sin\theta_i - \frac{2\pi}{\Lambda}\cos\theta_r \\ \left[\left(\frac{2\pi}{\lambda}\right)^2 - \left(K_d^X\right)^2 - \left(K_d^Y\right)^2\right]^{\frac{1}{2}} \end{bmatrix} = \frac{2\pi}{\lambda}\begin{bmatrix} -G\sin\theta_r \\ \sin\theta_i - G\cos\theta_r \\ [1 - G^2 - \sin^2\theta_i + 2G\sin\theta_i\cos\theta_R]^{\frac{1}{2}} \end{bmatrix} \tag{47}$$

The diffracted wave vector can then be expressed as:

$$\mathbf{K}_d = K_d^X \mathbf{e}_x + K_d^Y \mathbf{e}_y + K_d^Z \mathbf{e}_z \tag{48}$$

Subsequently, the scanning point trajectory from the local grating coordinate system can be calculated. It is evident that as the grating rotates, the scan trajectory follows a non-linear bow-like curve [29]. This results in a distorted LiDAR detection area as shown in Fig. 15d. Although the entire scan area is not a regular shape, the detection of each point is allocated to an absolute coordinate position, ensuring the accuracy of the image and dimensions. Finally, the point cloud data can easily be converted to the global coordinate system using a transform matrix $\mathbf{R}_{X'Y'Z'}$.

$$\mathbf{R}_{X'Y'Z'} = \begin{bmatrix} 1 & 0 & 0 \\ 0 & \cos\gamma & -\sin\gamma \\ 0 & \sin\gamma & \cos\gamma \end{bmatrix} \tag{49}$$

**Supplementary Note 11 | User identification (UID) code**

We adopt non-repeating necklace sequences when selecting UID codes to avoid sequence duplication issues caused by bit shifts from time delays, which inherently exist in an asynchronous CDMA system. This method ensures that each UID code is unique and cannot be generated by shifting any other UID code, ensuring linear independence in a set of vectors. The number of possible non-repeating necklace sequences is determined by the set number of bits as:

$$u(n) = \frac{1}{n}\sum_{\varepsilon|n} \phi(\varepsilon) \cdot 2^{n/\varepsilon} \tag{50}$$

where $u(n)$ represents the number of distinct binary sequence combinations of bit length $n$, $\phi(\varepsilon)$ is Euler's totient function, and $\varepsilon$ is a divisor of $n$. In our example in the main text, a bit number of 3 can generate four combinations: 111, 000, 101, and 010. This can also be verified by calculating:

$$u(3) = \frac{1}{3}(\phi(1) \cdot 2^3 + \phi(3) \cdot 2^1) = 4 \tag{51}$$

As mentioned in the main text, employing extra coding for UID code can generate more candidate sequences using less spreading length, thereby enriching the sequence pool. This can be understood in the following way. Consider a Gold sequence of order $m$, the number of chip periods is $N = 2^m - 1$ and the number of available orthogonal codes is $2^m + 1$. Next, when a UID code of bit length $n$ is added to the Gold sequence of order $m$, it increases the length of the sequence to $n \cdot (2^m - 1)$ chip periods, and at the same time, the number of unique sequence choices for identity-enabled CDMA ranging increases to $(2^m + 1) \cdot u(n)$. For simplicity, let us consider a UID bit length of order $l$ ($n = 2^l$); the total sequence length is then $n \cdot (2^m - 1) = 2^l \cdot (2^m - 1) < 2^{m+l} - 1$, which is shorter than the total chip period of a Gold sequence of order $m + l$. On the other hand, when using a direct sequence spreading without UID, the number of available orthogonal codes is $2^{m+l} + 1$ for a Gold sequence of order $m+l$. In summary, the number of candidate sequences for both methods can be calculated by:

$$UID: (2^m + 1) \cdot u(2^l) \tag{52}$$

$$Direct\ spread: 2^{m+l} + 1 \tag{53}$$

Using a 9th-order code, a comparison between the two methods is shown in the Fig. 17 and demonstrates that employing user code can lead to an increase in sequence candidates with a shorter total sequence length.

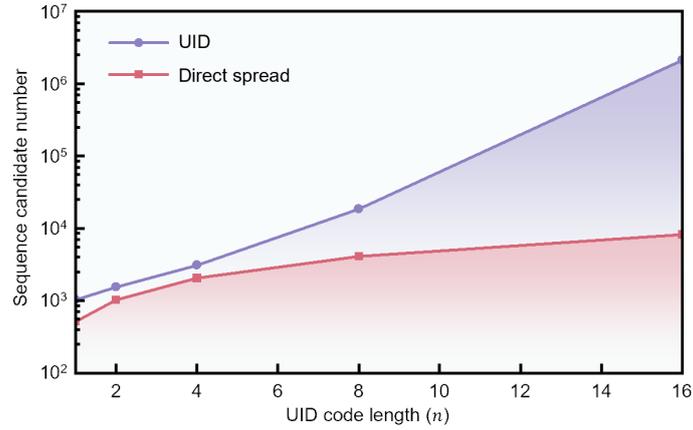

Fig. 17. Increment of sequences candidate number from 9th-order code with UID.

We conduct multiple experiments with UID to investigate various practical scenarios. Four unique UIDs can be generated from three bits. We simulate the scenario where four users use the same Gold sequence simultaneously but with different UID codes. Specifically, a common scenario involves different users being affected by interferences from other users at non-identical time delays. This situation is depicted in Main **Fig.4b**, where different distances influence signal strengths, thus resulting in different peak values. Clearly, the differentiation based on UID enables the identification of the four users. It is important to note that accurate distance measurement for each user is only possible when cross-correlating with its own synchronized Gold sequence code. Without synchronized codes from the other users, the decoded distances of other users appear randomly distributed along the relative time axis. This is perfectly fine, since the purpose of the UID is to identify and remove the incorrect ranging information created by the interferences from other users having the same spreading code.

**Supplementary Note 12 | Closed-loop control for near-far problem**

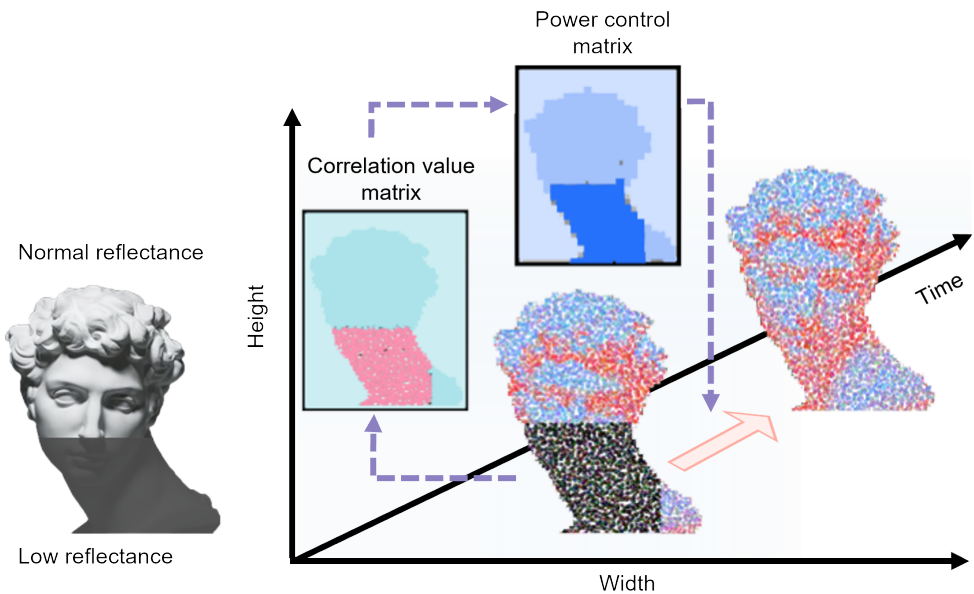

Fig. 18. Schematic of control loop to compensate for near-far problem.

Several factors contribute to the near-far problem beyond the effects of signal strength attenuation due to the varying distances of objects. Another potential cause is the different reflectivity of materials on the surfaces of the same object at similar distances. For instance, if certain materials have a higher absorption (hence low reflectivity) for the specific wavelength than adjacent areas, this can lead to a near-far problem as well; for example a partially painted sculpture in Fig. 18. This figure also illustrates the iterative process to address the near-far problem through a closed-loop control system. Initially, the $SNR_{XC}$ heatmap of the 3D point cloud reveals extraordinarily strong and very weak signal regions in the same data column due to the above-mentioned near-far problem. The corrected laser intensity is estimated based on the $SNR_{XC}$ values in these two regions, yielding a power control matrix. Subsequent frame scanning adjusts the laser intensity for different pixels or channels according to this power control matrix to mitigate the near-far problem and achieve satisfactory 3D point clouds. In practice, due to the use of a tunable laser source and the nonlinear amplification relationship of the EDFA, we simplify linear control by only adjusting the gain multiplier of the EDFA. Potential applications can then obtain a strict linear control relationship to feedback and compensate for this issue until all objects are distinguishable within the same frame. Regardless of any near-far problems caused by any situation, this matrix can be utilized for closed-loop control and thus achieve rapid data correction and real-time improvement.

**Supplementary Note 13 | Photos for experimental set-up**

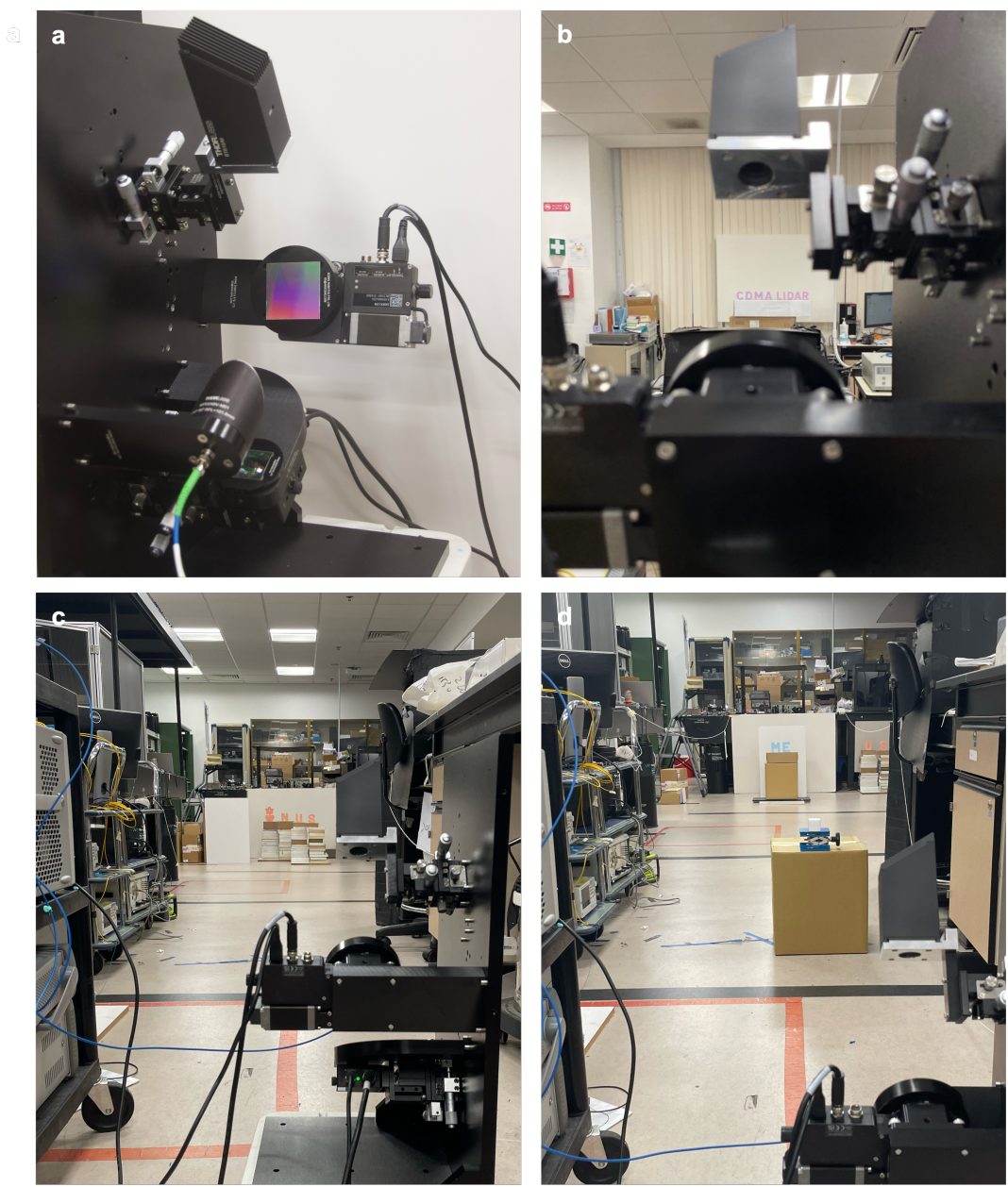

Fig. 19. Photos of the experimental setup: (a) Identity-enabled CDMA LiDAR prototype; setups for (b) 5 m scan, (c) 10 m scan, and (d) near-far problem.